\documentclass[10pt,journal,compsoc]{IEEEtran}
\usepackage{xurl}
\usepackage{amsfonts}
\usepackage{hyperref}
\usepackage{enumitem}
\usepackage[draft]{changes}
\usepackage[normalem]{ulem}
\usepackage{amsmath}
\usepackage{amssymb}
\usepackage{amsthm}
\usepackage{multirow}
\usepackage{makecell}

\ifCLASSOPTIONcompsoc
  \usepackage[nocompress]{cite}
\else
  \usepackage{cite}
\fi
\ifCLASSINFOpdf
\else
\fi
\begin{document}
%
% paper title
% Titles are generally capitalized except for words such as a, an, and, as,
% at, but, by, for, in, nor, of, on, or, the, to and up, which are usually
% not capitalized unless they are the first or last word of the title.
% Linebreaks \\ can be used within to get better formatting as desired.
% Do not put math or special symbols in the title.
\title{LNGate$^2$: Secure Bidirectional IoT Micro-payments using Bitcoin's Lightning Network and Threshold Cryptography*}

\author{Ahmet~Kurt,
        Kemal~Akkaya,
        Sabri~Yilmaz,
        Suat~Mercan,
        Omer~Shlomovits,
        and~Enes~Erdin~% <-this % stops a space
\IEEEcompsocitemizethanks{\IEEEcompsocthanksitem A. Kurt, K. Akkaya and S. Mercan are with the Department of Electrical and Computer Engineering, Florida International University, Miami, FL, 33174.
% note need leading \protect in front of \\ to get a newline within \thanks as
% \\ is fragile and will error, could use \hfil\break instead.
E-mail: \{akurt005, kakkaya, smercan\}@fiu.edu
\IEEEcompsocthanksitem S. Yilmaz is with the Department of Economics and Finance, Penn State Harrisburg. E-mail: suy22@psu.edu
\IEEEcompsocthanksitem O. Shlomovits is with Ingonyama. E-mail: omer.shlomovits@gmail.com
\IEEEcompsocthanksitem E. Erdin is with the Department of Computer Science and Engineering, University of Central Arkansas. E-mail: eerdin@uca.edu
\IEEEcompsocthanksitem *A preliminary version \cite{kurt2021lngate} of this work was published in the proceedings of the 14th ACM Conference on Security and Privacy in Wireless and Mobile Networks (ACM WiSec '21).}% <-this % stops an unwanted space

%\thanks{Manuscript received April 19, 2005; revised August 26, 2015.}
}

% note the % following the last \IEEEmembership and also \thanks - 
% these prevent an unwanted space from occurring between the last author name
% and the end of the author line. i.e., if you had this:
% 
% \author{....lastname \thanks{...} \thanks{...} }
%                     ^------------^------------^----Do not want these spaces!
%
% a space would be appended to the last name and could cause every name on that
% line to be shifted left slightly. This is one of those "LaTeX things". For
% instance, "\textbf{A} \textbf{B}" will typeset as "A B" not "AB". To get
% "AB" then you have to do: "\textbf{A}\textbf{B}"
% \thanks is no different in this regard, so shield the last } of each \thanks
% that ends a line with a % and do not let a space in before the next \thanks.
% Spaces after \IEEEmembership other than the last one are OK (and needed) as
% you are supposed to have spaces between the names. For what it is worth,
% this is a minor point as most people would not even notice if the said evil
% space somehow managed to creep in.

% The paper headers
\markboth{}%
{Shell \MakeLowercase{\textit{Kurt et al.}}: LNGate$^2$: Secure Bidirectional IoT Micro-payments using Bitcoin's Lightning Network and Threshold Cryptography}
% The only time the second header will appear is for the odd numbered pages
% after the title page when using the twoside option.
% 
% *** Note that you probably will NOT want to include the author's ***
% *** name in the headers of peer review papers.                   ***
% You can use \ifCLASSOPTIONpeerreview for conditional compilation here if
% you desire.

% The publisher's ID mark at the bottom of the page is less important with
% Computer Society journal papers as those publications place the marks
% outside of the main text columns and, therefore, unlike regular IEEE
% journals, the available text space is not reduced by their presence.
% If you want to put a publisher's ID mark on the page you can do it like
% this:
%\IEEEpubid{0000--0000/00\$00.00~\copyright~2015 IEEE}
% or like this to get the Computer Society new two part style.
%\IEEEpubid{\makebox[\columnwidth]{\hfill 0000--0000/00/\$00.00~\copyright~2015 IEEE}%
%\hspace{\columnsep}\makebox[\columnwidth]{Published by the IEEE Computer Society\hfill}}
% Remember, if you use this you must call \IEEEpubidadjcol in the second
% column for its text to clear the IEEEpubid mark (Computer Society jorunal
% papers don't need this extra clearance.)

% use for special paper notices
%\IEEEspecialpapernotice{(Invited Paper)}

% for Computer Society papers, we must declare the abstract and index terms
% PRIOR to the title within the \IEEEtitleabstractindextext IEEEtran
% command as these need to go into the title area created by \maketitle.
% As a general rule, do not put math, special symbols or citations
% in the abstract or keywords.
\IEEEtitleabstractindextext{%
\begin{abstract}
Bitcoin has emerged as a revolutionary payment system with its decentralized ledger concept; however it has significant problems such as high transaction fees and low throughput. Lightning Network (LN), which was introduced much later, solves most of these problems with an innovative concept called off-chain payments. With this advancement, Bitcoin has become an attractive venue to perform micro-payments which can also be adopted in many IoT applications (e.g., toll payments). Nevertheless, it is not feasible to host LN and Bitcoin on IoT devices due to the storage, memory, and processing restrictions. Therefore, in this paper, we propose a secure and efficient protocol that enables an IoT device to use LN's functions through an untrusted gateway node. Through this gateway which hosts the LN and Bitcoin nodes, the IoT device can open \& close LN channels and send \& receive LN payments. This delegation approach is powered by a threshold cryptography based scheme that requires the IoT device and the LN gateway to jointly perform all LN operations. Specifically, we propose thresholdizing LN's Bitcoin public and private keys as well as its public and private keys for the new channel states (i.e., commitment points). We prove with a game theoretical security analysis that the IoT device is secure against collusion attacks. We implemented the proposed protocol by changing LN's source code and thoroughly evaluated its performance using several Raspberry Pis. Our evaluation results show that the protocol; is fast, does not bring extra cost overhead, can be run on low data rate wireless networks, is scalable and has negligible energy consumption overhead. To the best of our knowledge, this is the first work that implemented threshold cryptography in LN.
\end{abstract}

% Note that keywords are not normally used for peerreview papers.
}

% make the title area
\maketitle

% To allow for easy dual compilation without having to reenter the
% abstract/keywords data, the \IEEEtitleabstractindextext text will
% not be used in maketitle, but will appear (i.e., to be "transported")
% here as \IEEEdisplaynontitleabstractindextext when the compsoc 
% or transmag modes are not selected <OR> if conference mode is selected 
% - because all conference papers position the abstract like regular
% papers do.
\IEEEdisplaynontitleabstractindextext
% \IEEEdisplaynontitleabstractindextext has no effect when using
% compsoc or transmag under a non-conference mode.

% For peer review papers, you can put extra information on the cover
% page as needed:
% \ifCLASSOPTIONpeerreview
% \begin{center} \bfseries EDICS Category: 3-BBND \end{center}
% \fi
%
% For peerreview papers, this IEEEtran command inserts a page break and
% creates the second title. It will be ignored for other modes.
\IEEEpeerreviewmaketitle

\IEEEraisesectionheading{\section{Introduction}\label{sec:introduction}}
% Computer Society journal (but not conference!) papers do something unusual
% with the very first section heading (almost always called "Introduction").
% They place it ABOVE the main text! IEEEtran.cls does not automatically do
% this for you, but you can achieve this effect with the provided
% \IEEEraisesectionheading{} command. Note the need to keep any \label that
% is to refer to the section immediately after \section in the above as
% \IEEEraisesectionheading puts \section within a raised box.

% The very first letter is a 2 line initial drop letter followed
% by the rest of the first word in caps (small caps for compsoc).
% 
% form to use if the first word consists of a single letter:
% \IEEEPARstart{A}{demo} file is ....
% 
% form to use if you need the single drop letter followed by
% normal text (unknown if ever used by the IEEE):
% \IEEEPARstart{A}{}demo file is ....
% 
% Some journals put the first two words in caps:
% \IEEEPARstart{T}{his demo} file is ....
% 
% Here we have the typical use of a "T" for an initial drop letter
% and "HIS" in caps to complete the first word.
\IEEEPARstart{T}{he} last decade has witnessed a fast adoption of IoT in numerous domains as it provides great opportunities and convenience \cite{gubbi2013internet}. These devices have typically been utilized for continuous data collection as they are equipped with various sensors. The improvement in their computational capabilities has enabled applications where an IoT device may pay/get paid for received/provided services. For instance, a vehicle passing through a toll gate can make a payment using its on-board unit which acts as an IoT device to communicate with the toll infrastructure \cite{pavsalic2016vehicle}. Other potential \textit{micro-payment} applications in this context include electric vehicle charging, parking payments, sensor data trading etc. \cite{mercan2021cryptocurrency, kurt2020lnbot}.

Since these micro-payments are generally device-to-device where no human intervention is involved, they should be automated. Although it might be possible to link these devices to traditional payment systems such as credit cards, we argue that as seen with some existing applications \cite{radenkovic2023digital}, some users prefer cryptocurrency payments over traditional methods due to the privacy and anonymity advantages or simply due to personal choice and thus offering a cryptocurrency-based payment solution for IoT transactions is needed. Out of all cryptocurrencies, Bitcoin \cite{nakamoto2008bitcoin} has the biggest potential to be used for IoT micro-payments as it is the most commonly used cryptocurrency and dominate the market in terms of market capitalization. Specifically, out of all cryptocurrencies, its market cap is at around 50\%\footnote{\url{https://coinmarketcap.com/}}. Recent studies show that it became a viable option compared to credit card payments, especially in Asia \cite{asiacrypto}. Its current user base is more than 150 million\footnote{\url{https://blog.chainalysis.com/reports/bitcoin-addresses/}}. This means that regardless of the debate around Bitcoin, there is a growing community out there which will continue to use it and thus, there will always be a need to cater to their needs in terms of research and development.

However; long confirmation times, low throughput, and high transaction fees have been the main concern in mainstream cryptocurrencies which limits their scalability \cite{chauhan2018blockchain} as well as inhibits their adoption for micro-payment scenarios. Thus, payment channel network (PCN) idea was proposed which addresses these problems by utilizing off-chain transactions using a second-layer on top of the blockchain \cite{decker2015fast}. As an example, Lightning Network (LN) \cite{poon2016bitcoin} has been introduced as a PCN solution for Bitcoin. LN currently exceeds 16,000 nodes since its creation five years ago. In addition to addressing the scalability, latency and cost issues, LN also provides flexibility in terms of its reach. Specifically, as long as one has at least one LN channel, s/he can make payments to any other LN nodes using the existing channels in the network as long as there is a multi-hop route. Despite these advantages, LN still cannot be run on most of the IoT devices because of the computation, communication, and storage requirements \cite{lniotproblem}. Specifically, using LN requires running an LN node along with a full Bitcoin node which in total, occupies more than 473 GB\footnote{\url{https://ycharts.com/indicators/bitcoin_blockchain_size}} of storage at the time of writing this paper. Additionally, the devices need robust Internet connection and relatively high computation power to receive and verify the new Bitcoin blocks.

Therefore, a lightweight solution is needed to enable resource-constrained IoT devices to use LN for micro-payments. We are specifically focusing on Bitcoin's LN and not Ethereum based PCNs because LN is currently the most widely used cryptocurrency payment channel network. To this end, in this paper, we propose a threshold cryptography-based protocol where an IoT device can perform LN operations through an \textit{untrusted LN gateway} that hosts the full LN and Bitcoin nodes. With this integration, the IoT device can 1) open LN channels, 2) send LN payments, 3) receive LN payments, and 4) close LN channels. The LN gateway is incentivized to provide this payment service by charging service fees for IoT device's payments.

In our proposed protocol, we utilize (2,2)-threshold cryptography to enable IoT device to control aforementioned LN operations without having to trust the LN gateway. The IoT device and the LN gateway cooperatively authorize LN operations by incorporating their threshold secret shares into signing and key generation operations. In this way, the LN gateway cannot spend IoT device's funds in the channel without the IoT device joining in a (2,2)-threshold signing. We propose changes to LN's peer channel protocol and commitment transactions. With these modifications, LN channels can also accompany IoT device's funds while maintaining the existing security features of LN such as protection from cheating attempts from channel peers. We prove that our proposed protocol is secure against collusion attacks using a game theoretical security analysis.

To assess the effectiveness and overhead of the proposed protocol, we implemented it by changing the source code of one of the main LN implementations. In our test setup, we used several Raspberry Pis as IoT devices and a desktop computer as the LN gateway. We demonstrated that the proposed protocol 1) has negligible communication and computational delays thus enables timely payments; 2) is scalable for increasing number of IoT devices and payments; 3) can be run on networks with low bandwidth (data rate); and 4) associated energy consumption and monetary costs of using the protocol for IoT devices are negligible.

\vspace{1mm}
\textbf{Contributions}: Our contributions in this work are as follows:
% \vspace{-1mm}
\begin{itemize}[leftmargin=*]

    \item We propose LNGate$^2$, a secure and lightweight protocol that enables resource-constrained IoT devices to open/close LN channels, send/receive LN payments through an untrusted gateway node. 
    
    \item We utilize (2,2)-threshold cryptography so that the IoT device does not have to get involved in Bitcoin or LN operations, but only in transaction signing and key generation. We propose to thresholdize LN's Bitcoin public/private keys and public/private keys of new channel states (i.e., commitment points) for a secure 2-party threshold LN node. 
    
    \item We used game theory to analyze the security of the protocol and prove that it is secure against collusion attacks.
    
    \item We implemented LNGate$^2$ by changing LN's source code. LN's Bitcoin public and private keys were thresholdized. Our code is publicly available at our GitHub page: \url{https://github.com/startimeahmet/lightning}.

\end{itemize}

The rest of the paper is structured as follows. In Section \ref{sec:RelatedWork}, we discuss the related work. Section \ref{sec:background} describes the LN, threshold cryptography and game theory preliminaries. System and threat model are given in Section \ref{sec:systemmodel}. The proposed protocol is explained in detail in Section \ref{sec:protocol}. Section \ref{sec:security} presents the security analysis of the protocol. Detailed performance evaluation of the proposed protocol is given in Section \ref{sec:evaluation}. Limitations of the approach is mentioned at Section \ref{sec:limitations}. Finally, we conclude the paper in Section \ref{sec:conclusion}.

\section{Related Work}
\label{sec:RelatedWork}

Hannon and Jin \cite{hannon2019bitcoin} propose a protocol based on LN to give the IoT devices the ability to transact. They propose using two third parties that they named \textit{IoT payment gateway} and \textit{watchdog}. However, their approach has a fundamental flaw. They assumed that the IoT device can open payment channels to the gateway and this process is not explained. This is indeed the exact problem we are trying to solve. IoT devices do not have the computational resources to open and maintain LN payment channels. Therefore, their assumption is not feasible and our work is critical in this sense to fill this gap. 

The authors of \cite{robert2020enhanced} proposed IoTBnB which is a digital IoT marketplace that utilizes LN payments for data trading. In their approach, an \textit{LN module} which hosts the Bitcoin and LN nodes is used to send users' payments. In contrast to our work, this approach is focusing on integrating LN into an existing IoT marketplace. Thus, the individual devices that are not part of such marketplaces are not considered. Additionally, the authors' LN framework relies on Bitcoin wallets held by the ecosystem itself which raises security and privacy concerns. In our approach, IoT devices do not share the ownership of their Bitcoins with a third party.

A work focusing on Ethereum micro-payments rather than Bitcoin was proposed by Pouraghily and Wolf \cite{pouraghily2019lightweight}. It is a ticket-based verification protocol to enable low-end IoT devices to exchange money and data inside an IoT ecosystem. However, this approach has major problems: The joint account opened with a partner device raises security concerns as the details of it are not provided. Additionally, the approach was compared with $\mu$Raiden \cite{raiden} whose development stopped more than 4 years ago. In contrast to this work, we targeted Bitcoin's LN as it is actively being developed and dominating the market.

A recent work by Rebello et al. \cite{rebello2022securing} proposed a hybrid PCN architecture for wireless resource constrained devices to be able to access and use PCNs. However, their solution does not work for devices that stay offline for long periods of time. Additionally, authors assume that resource-constrained devices can run light nodes and establish payment channels. This assumption cannot be generalized to all the IoT devices. Our approach does not require IoT devices to stay online except for the time they perform the LN operations. They also do not need to run light nodes.

Another recent work by Wang et al. \cite{wang2021hyperchannel} introduced HyperChannel which
utilizes a group of Intel Software Guard Extensions (SGXs) that are run by selfish service nodes to execute the transactions. The approach includes an entity called \textit{client emergency enclave}. This is an extra burden on the IoT devices since each device has to own an emergency enclave to protect themselves in case of service node shutdowns. In our approach, we only require IoT devices to perform signing.

Profentzas et al. \cite{profentzas2020tinyevm} proposed TinyEVM to enable IoT devices to perform micro-payments. The authors' method involves running a modified version of Ethereum virtual machine on the IoT devices. In contrast, in our approach, IoT devices only generate signatures which is not a resource-intensive operation. The work by Li et al. \cite{li2020data} focuses on designing a PCN-based smart contract for IoT data transactions. However, they do not seem to discuss the costs associated with their protocol and the routing performance of the protocol drops significantly when there are malicious nodes in the network. A slightly different work by Tapas et al. \cite{tapas2020p4uiot} proposed utilizing LN in the context of patch update delivery to the IoT devices where they claim rewards through LN. However, unlike our work, the IoT devices are assumed to be able to connect to the blockchain through light clients or third party full nodes on the network which requires a degree of trust.

There are also some implementation efforts for creating lighter versions of LN such as Neutrino \cite{neutrino} light client and Phoenix \cite{phoenix} mobile wallet. The problem with these software is that they are not specifically designed for IoT devices thus most IoT devices cannot run them. Thus, we opt for a solution that covers a wider range of IoT devices and applications.

In addition to these works, we acknowledge that there are also many works that offer cash payments from IoT devices that do not include cryptocurrencies. These are typically bank-based, card-based and digital cash based payment options \cite{wang2020untraceable, bojjagani2022secure}. However, as mentioned before, our goal is not to compete with these solutions. Our work is just a reliable and convenient alternative for those who prefer cryptocurrency payments that rely on blockchains.

Finally, we would like to note that this work is an extension of our conference paper \cite{kurt2021lngate} with a lot of additions and improvements as follows: 1) The proposed protocol in the conference version only supports unidirectional payments. In this paper, we proposed a new protocol where the IoT device can also receive payments in addition to sending; 2) In the conference version, the security of the protocol against collusion attacks was not analyzed formally. In this journal version, we used game theory to formally analyze the collusion attacks by considering both unidirectional and bidirectional payment cases; 3) The evaluation section in the conference version only has limited WiFi experiments. In this paper, we present Bluetooth Low Energy (BLE) experiments in addition to the WiFi. We also present completely new scalability and energy consumption experiments. Overall, we offer more experiments, analysis and details; 4) Overall, almost all the sections from the conference version is either rewritten, modified significantly or extended.

\section{Background}
\label{sec:background}
This section provides background on LN, its underlying mechanisms, threshold cryptography and game theory as a preliminary to our proposed approach. 

\subsection{Lightning Network Preliminaries}
\label{sec:preliminaries}

LN was introduced in 2015 in a draft technical whitepaper \cite{poon2016bitcoin} and later was implemented and deployed onto Bitcoin Mainnet by Lightning Labs \cite{lnlaunch}. It runs on top of the Bitcoin blockchain as a \textit{second-layer} peer-to-peer distributed PCN and aims to address the scalability problem of Bitcoin. It enables opening secure payment channels among users to perform instant and cheap Bitcoin transfers through multi-hop routes within the network by utilizing Bitcoin's smart contract capability \cite{lande2018sok}. The number of users using LN has grown significantly since its creation. At the time of writing this paper, LN incorporates 16,450 nodes and 74,625 channels which hold 5,438 BTC in total (worth around 160 million USD)\footnote{\url{https://1ml.com/}}.

The rest of this section explains technical concepts about LN such as \textit{funding transaction}, \textit{commitment transaction}, \textit{Hash Time Locked Contract (HTLC)}, \textit{revoked state}, \textit{key send \& invoice payments} and \textit{Basis of Lightning Technology (BOLT)} which are essential to understand our protocol. The explanations are based on an example case where Alice opens an LN payment channel to Bob and wishes to transact with him.

\vspace{1mm}

\noindent \textbf{Funding Transaction}: When Alice would like to open a channel to Bob, this is done via an on-chain Bitcoin transaction called the \textit{funding transaction}. Generally the channel is funded by the party initiating the process, but dual-funding where both channel parties commit funds to the channel is also possible. With the funding process, the capacity of the channel is determined as well. For example, if Alice funds the channel with 5 Bitcoins (BTC), she can send payments to Bob until her 10 BTC in the channel are exhausted. 

The reverse of the funding transaction is called the \textit{closing transaction} which is used to close an LN channel. It is also an on-chain transaction which needs to be broadcast to the Bitcoin network.

\begin{figure}[h]
    \centering
    % \vspace{-2mm}
    \includegraphics[width=0.60\linewidth]{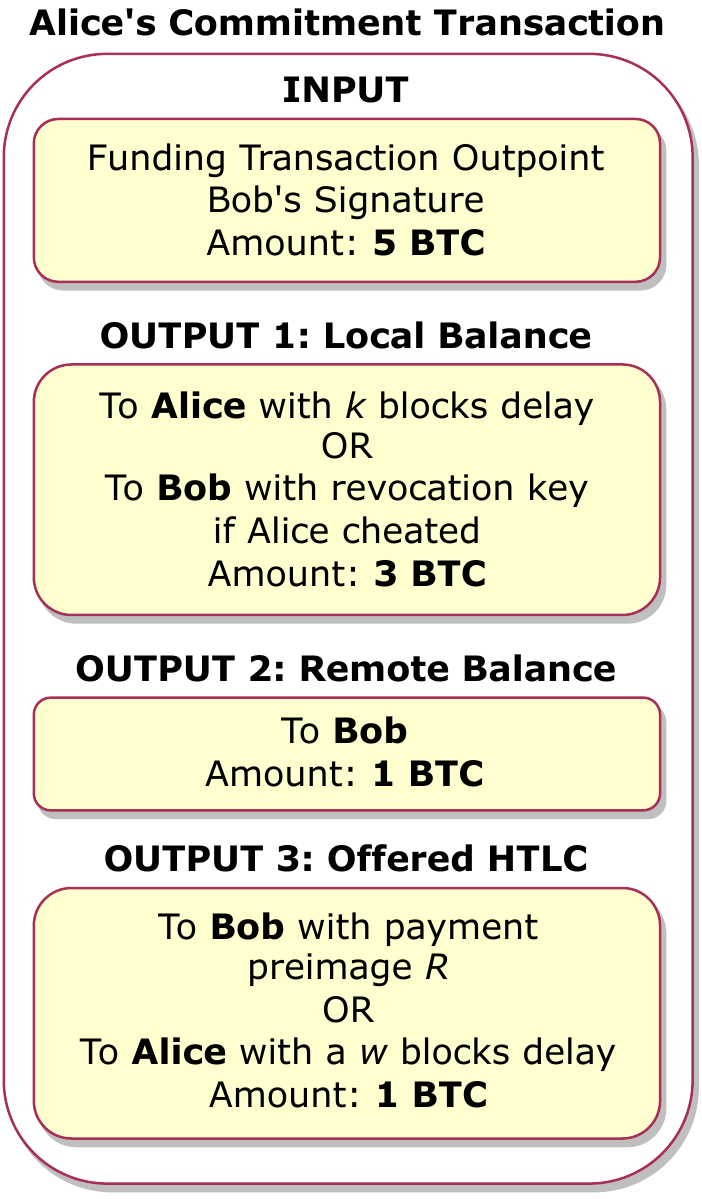}
    \vspace{-4mm}
    \caption{An illustration of a commitment transaction stored at Alice.}
    \label{fig:Alice_commitment}
    \vspace{-7mm}
\end{figure}

\vspace{1mm}

\noindent \textbf{Commitment Transaction}: Once the channel is used for sending a payment, the balances of Alice and Bob in the channel will change. Since the LN payments are off-chain meaning they are not recorded on the Bitcoin blockchain, another type of Bitcoin transaction which will keep track of the channel balances is needed. This is done by the \textit{commitment} system of LN. A simple LN payment requires two separate commitment rounds: one when the payment is offered and one when it is fulfilled. Each commitment round requires both peers to sign a \textit{commitment transaction}. Here, the commitment transactions are what actually hold the channel balance information and incoming/outgoing payments. They are specially crafted for LN and symmetrical for channel peers.

Inputs to Alice's commitment transaction is the funding transaction outpoint and Bob's signature. Outpoint is the combination of the transaction output and its output index. A typical commitment transaction has three main outputs. For Alice's version, the first output is for her balance, the second output is for Bob's and the third would be for an incoming/outgoing payment. These outputs are illustrated in Fig. \ref{fig:Alice_commitment}.

\noindent \textbf{Hash Time Locked Contract (HTLC)}: LN payments are constructed using a special type of transaction called Hash Time Locked Contract (HTLC). In this scheme, when Alice wants to send a 1 BTC payment to Bob, she asks Bob to generate a secret called \textit{preimage}. Bob hashes the preimage and sends the hash to Alice. Alice includes the hash in the HTLC and sends the HTLC to Bob. Upon receiving Alice's HTLC, Bob reveals the preimage to Alice to receive the 1 BTC payment. This acts as a proof that Bob is the intended recipient. If for some reason, Bob cannot reveal the preimage on time to claim the HTLC, Alice gets the 1 BTC back after \textit{w} new blocks are mined. Output 3 in Fig. \ref{fig:Alice_commitment} illustrates this HTLC payment.

\vspace{1mm}

\noindent \textbf{Revoked State}: Alice's balance in her commitment transaction is conditional such that; if she closes the channel, she has to wait \textit{k} number of blocks before she can redeem her funds on-chain. This is to protect Bob from a possible cheating attempt by Alice in which she uses an old (\textit{revoked}) channel state to close the channel. Since Alice has a record of all the old channel states, it might be tempting for her to broadcast an old state in which she has more funds. But, if Bob sees that the channel was closed using an old state, he will punish Alice by taking all her funds in the channel. But to do that, he has to be online and take action before the broadcast transaction reaches the depth \textit{k} on the blockchain. Note that, when Alice closes the channel, Bob can redeem his funds immediately unlike Alice.

\noindent \textbf{Key Send \& Invoice Payments}: 
LN payments are sent using \textit{invoices} which are basically long strings consisting of characters and numbers\footnote{\url{https://github.com/lightning/bolts/blob/master/11-payment-encoding.md}}. They are encoded in a specific way to include all the information required to send the payment such as the destination public key, payment amount, expiration date, signature etc. The recipient of the payment prepares the invoice and sends it to the payer. A more convenient payment method in LN which does not involve such preparation is called \textit{key send}\footnote{\url{https://github.com/lightningnetwork/lnd/pull/3795}}. With this method, senders do not need to contact the recipients or have an invoice first to send payments. 

Note that as long as there is at least one channel to LN, one may be able to reach all other LN nodes since LN enables multi-hop payment forwarding. This means one does not need to open an LN channel with every node it wants to exchange payments.

\vspace{1mm}

\noindent \textbf{Basis of Lightning Technology (BOLT)}: BOLT specifications describe LN's layer-2 protocol for secure off-chain Bitcoin payments. In order to implement our proposed protocol, we made modifications on BOLT \#2 which is LN's peer protocol for channel management. BOLT \#2 has three phases which are \textit{channel establishment}, \textit{normal operation} of the channel, and \textit{channel closing}. Using this protocol, LN nodes talk to each other for channel related operations. The full details of the protocol can be found at \cite{bolt2}.

\subsection{Threshold Cryptography}
\label{sec:thresholdbackground}

\textit{Threshold cryptography} \cite{desmedt1994threshold} deals with cryptographic operations where more than one party is involved. The idea of sharing a cryptographic secret among a number of parties was proposed by Shamir \cite{shamir1979share}. In a threshold scheme, a secret is shared among $n$ parties and a threshold $t$ is defined such that, no group of $t-1$ can learn anything about the secret. Such setup is defined as $(t, n)$-threshold scheme. We utilize (2,2)-threshold cryptography in our proposed protocol for cooperative signing and key generation. The (2,2)-threshold schemes we used in this work are the same as the ones we used in our preliminary work \cite{kurt2021lngate}. We just provide the details of the Elliptic Curve Digital Signature Algorithm (ECDSA) which is used by Bitcoin for its signing operations.

\vspace{1mm}

\noindent \textbf{ECDSA Signature Scheme}: ECDSA signature scheme takes an input message $m$ and a private key $x$ and produces a pair of integers $(r, s)$ as output. The steps of the algorithm are as follows:

\begin{enumerate}[leftmargin=*]
    \item Hash $h = H(m)$ of the message is calculated. $H$ is a hash function (i.e., SHA-256).
    \item A secure random integer $k$ between $[1, n-1]$ is generated.
    \item A random point $(x, y) = k \cdot G$ is calculated. Then, $r = x$ $(mod$ $n)$ is computed.
    \item Signature proof $s = k^{-1} \cdot (h + r \cdot x)$ $(mod$ $n)$ is calculated.
    \item $(r,s)$ is returned which is the ECDSA signature.
\end{enumerate}

\subsection{Game Theory Background}
In this paper, the concept of \textit{subgame perfect equilibrium} (SPE) \cite{moore1988subgame} plays an important role to identify the solution. A (proper) subgame is a subset of the tree structure of the actual game. When tree form is used in a game, sequentiality can be applied rather than simultaneity. That is, in a two-player game, one player moves first and the other decides how to play after observing the first player’s action. In this setting, it is assumed that the players are all sequentially rational, meaning that each player systematically and purposefully performs his/her best to achieve the objective. Most of the time, the objective for a player in an economic setting is to maximize profit. One should also note that profit maximization requires the cost minimization at the same time by its definition. 

To find the solution of a game with a similar setting (i.e., SPE), the concept called \textit{backward induction} \cite{aumann1995backward} can be used. Applying the backward induction procedure requires investigating the game from the end to the start. Particularly, at each decision node, any action that results in a smaller payoff for the corresponding player is eliminated. Incorporating sequential rationality, a strategy profile is SPE if it specifies a \textit{Nash equilibrium} \cite{myerson1978refinements} in every subgame of the original game. A Nash equilibrium is the optimal solution which yields no incentive for any player in the game to deviate once it is achieved.

\begin{figure}[h]
    \centering
    \vspace{-1mm}
    \includegraphics[width=\linewidth]{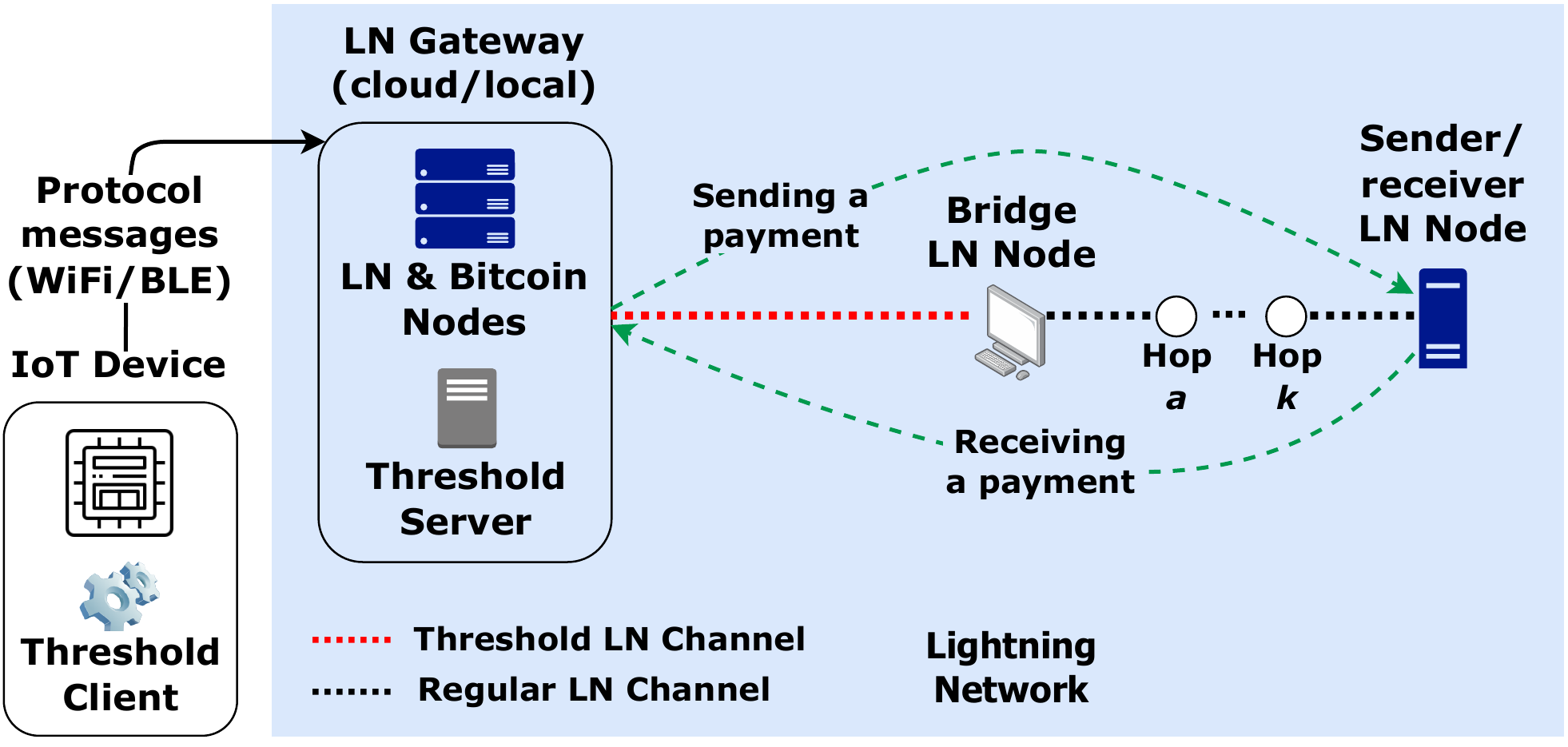}
    \vspace{-7mm}
    \caption{Illustration of the system model.}
    \label{fig:system_model}
    \vspace{-6mm}
\end{figure}

\begin{figure}[t]
    \centering
    % \vspace{-2mm}
    \includegraphics[width=\linewidth]{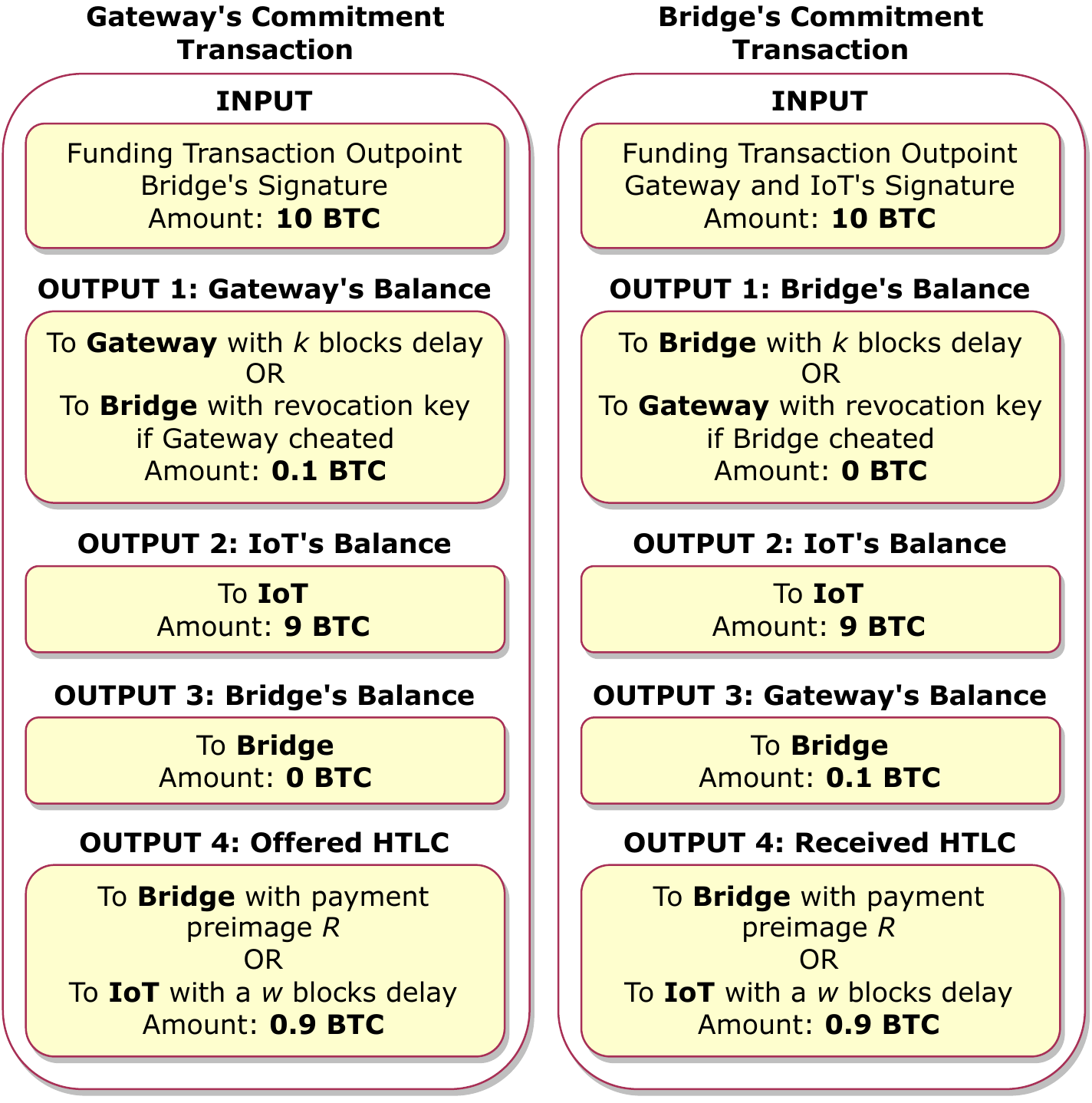}
    \vspace{-7mm}
    \caption{Depiction of the proposed commitment transactions for the LN gateway and the bridge LN node. These commitment transactions are generated after the following operations: 1) A channel with 10 BTC capacity was opened, 2) IoT requested sending 1 BTC to a destination, 3) Gateway charged the IoT a service fee of 0.1 BTC for this payment (the fee in real life would be much less).}
    \vspace{-2mm}
    \label{fig:gatewaybridgecommitment}
\end{figure}

\section{System \& Threat Model}
\label{sec:systemmodel}

\subsection{System Model}
There are four main entities in our system which are \textbf{IoT device}, \textbf{LN gateway}, \textbf{Bridge LN node}, and \textbf{Sender/receiver LN node} as shown in Fig. \ref{fig:system_model}. We also show other tools and intermediary devices; \textbf{Threshold client}, \textbf{LN gateway's LN and Bitcoin nodes}, \textbf{Threshold server}. IoT device wants to pay the receiver LN node for the goods/services. IoT device can also receive payments if needed such as a refund or a payment from another LN node (sender LN node). In other words, payments are bidirectional meaning that the IoT device can both send and receive payments. We assume that the owner of the IoT device also operates the device for these transactions. Any third party operation of the IoT devices is also possible but this may raise business privacy issues among the owner and the operator which is beyond the scope of this work. The LN gateway can be hosted on the cloud or locally depending on the use case scenario. It provides services to the IoT device by running the required full Bitcoin and LN nodes and is \textit{incentivized by the fees the IoT device pays in return}. The LN gateway also runs a threshold server that communicates with the threshold client installed at the IoT device when required. This client/server setup enables the 2-party threshold ECDSA operations. Bridge LN node is the node to which the LN gateway opens a channel when requested by the IoT device. Through the bridge LN node, IoT device's payments are either routed to a receiver LN node specified by the IoT device or delivered to the IoT device from a sender LN node on the Internet. 

We assume that the IoT device and the LN gateway do not go offline in the middle of a process such as sending or receiving a payment. IoT device can be offline for the rest of the time.

\subsection{Threat Model}
\label{sec:threatmodel}

We assume that the communication between the IoT device and the LN gateway is secured using TLS-like mechanisms. Based on our system model and application, possible adversaries to the system are the LN gateway and the bridge LN node which are assumed to be malicious. Therefore, we consider the attacks that would potentially be performed by these actors as well as the ones related to the specific processes of our payment application for the IoT devices. For instance, we assume that, both of these actors can be selfish in the sense that they can send old channel states to the Bitcoin blockchain in an attempt to cheat. We also assume that the LN gateway and the bridge LN node can collude with each other for deceiving the IoT devices. Furthermore, these nodes can also deviate from the proposed protocol descriptions to make monetary benefits.

Note that there may be also external attacks to Bitcoin's consensus mechanism and transactions independent from our approach i.e., not specific to our application. For instance, 51\% attack enables external adversaries to gain control of the blockchain \cite{saad2019overview}. Similarly, double spending attack tries to  enable spending of the same currency at least two times \cite{karame2012double}. There are also attacks to LN by congesting the channels or the layer-1 \cite{MizrahiCongestion21,sguanci2022mass}. Since mitigations to these known attacks are already analyzed in previous studies \cite{sayeed2019assessing, nicolas2020blockchain} or in the same papers where the attacks are proposed (e.g., for LN), we assume that our protocol will not be impacted from these attacks. Based on these assumptions, we consider the following attacks to our system:

\begin{itemize}[leftmargin=*]
    
    \item \textbf{Threat 1: Collusion Attacks:} The LN gateway and the bridge LN node can collude with each other to steal money from the IoT device.
     
    \vspace{1mm}
    
    \item \textbf{Threat 2: Stealing IoT Device's Funds:} The LN gateway can steal IoT device's funds that are committed to the channel by 1) sending them to other LN nodes; 2) broadcasting revoked states and; 3) colluding with the bridge LN node.

    \vspace{1mm}

    \item \textbf{Threat 3: Ransom Attacks:} The LN gateway can deviate from the protocol after opening a channel for the IoT device and not execute IoT device's requests (i.e., uncooperative LN gateway). Then, it can ask the IoT device to pay a ransom before executing the payment sending/receiving or channel closing operations.

    \vspace{1mm}
    
\end{itemize}

\section{Proposed Protocol Details}
\label{sec:protocol}

This section explains the details of the protocol that includes the channel opening, sending a payment, receiving a payment and channel closing. As mentioned in Section \ref{sec:preliminaries}, we propose modifications to LN's BOLT \#2 which are shown throughout the protocol descriptions. More importantly, LN's signing mechanism is modified and replaced with a (2,2)-threshold scheme that is utilized by the IoT device and the LN gateway. In addition to that, we propose changes to LN's commitment transactions to accompany IoT device's funds in the channel which is explained next.

\subsection{Modifications to LN's Commitment Transactions}
\label{sec:LN_modified}
Introduction of the IoT device requires some modifications to LN's commitment transactions as there are now 3 channel parties instead of 2. Since each channel party has a separate balance in the channel, they have to have an output in the commitment transactions reflecting their balance. An illustration of LN's original commitment transaction was given in Fig. \ref{fig:Alice_commitment}. 

Our proposed modified version of LN's commitment transactions are shown in Fig. \ref{fig:gatewaybridgecommitment}. As can be seen, there is an extra output for the IoT device in both versions of the commitment transaction. This output is not time-locked nor conditional unlike other outputs as the IoT device cannot be punished because of cheating attempts from other channel parties. This essentially protects IoT device's funds in the channel. Apart from that, the LN gateway's and the bridge LN node's outputs are regular time-locked outputs and are spendable by the counterparty in case of a cheating attempt.

\subsection{Channel Opening Process}
\label{sec:channelopening}

The IoT device is not able to open a channel by itself as it does not have access to LN nor Bitcoin network. Therefore, we enable the IoT device to securely initiate the channel opening process through the LN gateway and jointly generate signatures with it using the (2,2)-threshold scheme. This means that the LN's current channel opening protocol needs to be modified according to our needs. 

All the steps for the channel opening protocol which includes the default LN messages and our additions are depicted in Fig. \ref{fig:openchannel}. We explain the protocol step by step below:

\begin{figure}[h]
    \centering
    \vspace{-1mm}
    \includegraphics[width=\linewidth]{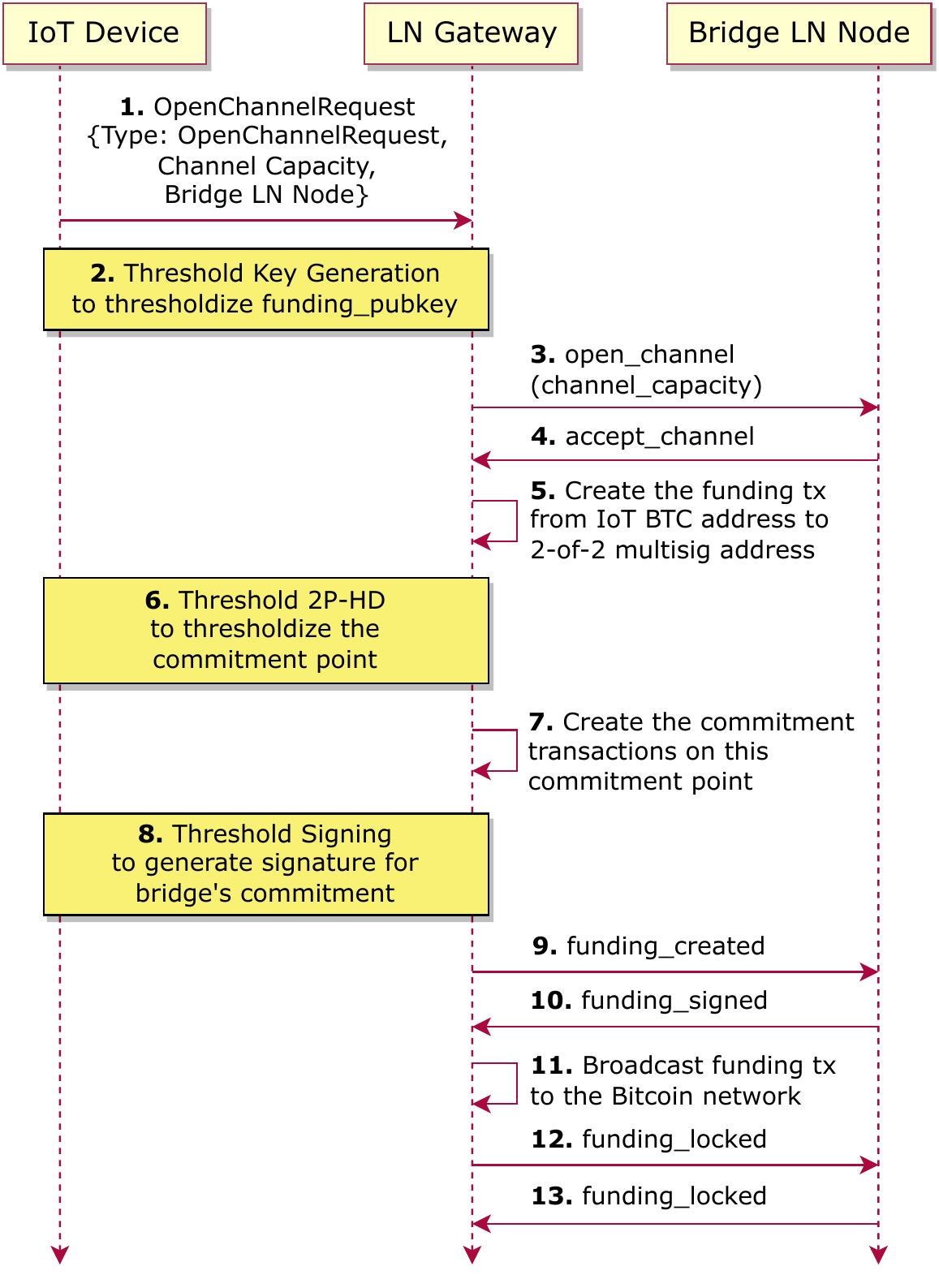}
    \vspace{-9mm}
    \caption{Protocol steps for opening a channel.}
    \label{fig:openchannel}
    \vspace{-2mm}
\end{figure}

\begin{itemize}[leftmargin=*]

\item \textbf{IoT Channel Opening Request}: IoT sends an \textit{OpenChannelRequest} message (Message \#1 in Fig. \ref{fig:openchannel}) to the LN gateway to request a payment channel to be opened to a bridge LN node. This message has the following fields: \textit{Type: OpenChannelRequest, Channel Capacity, Bridge LN Node}. \textit{Channel Capacity} is specified by the IoT device and this amount of Bitcoin is taken from IoT device's Bitcoin address as will be explained in the next steps. Here, we opt to let the IoT device choose the \textit{bridge LN node} since letting the LN gateway choose the bridge LN node might not be secure as will be shown in Section \ref{sec:gametheory}. IoT device can make this choice by accessing API services that provide public LN information\footnote{e.g., \url{https://amboss.space/}}.

\item \textbf{Channel Opening Initiation}: Upon receiving the request from the IoT device, the LN gateway initiates the channel opening process by connecting to the bridge LN node specified by the IoT device. Before initiating the channel opening process by sending an \textit{open\_channel} message, we propose the LN gateway to perform a (2,2)-threshold key generation with the IoT device to \textbf{thresholdize} its \textit{funding\_pubkey} (\#2 in Fig. \ref{fig:openchannel}). The \textit{funding\_pubkey} is a Bitcoin public key and both channel parties have their own. This process replaces the LN gateway's Bitcoin public and private keys with a threshold public/private key pair that is jointly computed between the IoT device and the LN gateway. In this way, the LN gateway cannot spend the funds the IoT device is committing to the channel without the IoT device's authorization. After this step, the LN gateway sends the \textit{open\_channel} message (\#3 in Fig. \ref{fig:openchannel}) to the bridge LN node which includes the thresholdized \textit{funding\_pubkey}. The \textit{channel capacity} specified by the IoT device is also sent with this message. After this step, the bridge LN node responds with an \textit{accept\_channel} message (\#4 in Fig. \ref{fig:openchannel}) to acknowledge the channel opening request of the LN gateway. Note that, \textit{open\_channel} and \textit{accept\_channel} are default BOLT \#2 messages.

\item \textbf{Creating the Transactions}: Now that the channel parameters are agreed on, the LN gateway can create a funding transaction from IoT device's Bitcoin address to the 2-of-2 multisignature address of the channel. Since the input to the funding transaction is from the IoT device's BTC address, IoT device can also pay the on-chain fee for this transaction\footnote{Custom funding transactions like this can be constructed by first creating a partially signed Bitcoin transaction (PSBT), then adding the inputs externally and finalizing it in LN. See: \url{https://lightning.readthedocs.io/lightning-openchannel_init.7.html}}. At this step, the LN gateway also creates the commitment transactions for itself and the bridge LN node (\#7 in Fig. \ref{fig:openchannel}). Here, we propose to compute the first \textit{commitment point} jointly between the IoT device and the LN gateway. Commitment points are used to derive revocation keys and they are unique for each channel state. Thus, thresholdizing the commitment points prevents the IoT device and the LN gateway from single-handedly revealing the revocation key before the channel state is updated. For this, we propose using a (2,2)-threshold child key derivation process (\#6 in Fig. \ref{fig:openchannel}) also known as, 2P-HD \cite{gothamcitywhitepaper}. 2P-HD allows the derivation of child keys from the master key that was already generated with (2,2)-threshold key generation earlier (\#2 in Fig. \ref{fig:openchannel}). We propose using 2P-HD over (2,2)-threshold key generation because of its efficiency.

\item \textbf{Exchanging Signatures}: Now, the LN gateway needs to send the signature for bridge LN node's version of the commitment transaction to the bridge LN node. For this, the LN gateway and the IoT device jointly generate the signature in a (2,2)-threshold signing (\#8 in Fig. \ref{fig:openchannel}). After signing is done, the LN gateway sends the signature to the bridge LN node along with the outpoint of the funding transaction in a \textit{funding\_created} message (\#9 in Fig. \ref{fig:openchannel}). Learning the funding outpoint, the bridge LN node is now able to generate the signature for the LN gateway's version of the commitment transaction and sends it over to the LN gateway in \textit{funding\_signed} message (\#10 in Fig. \ref{fig:openchannel}).

\item \textbf{Broadcasting the Transaction to the Bitcoin Network}: After the LN gateway receives the \textit{funding\_signed} message from the bridge LN node, it must broadcast the funding transaction to the Bitcoin network (\#11 in Fig. \ref{fig:openchannel}). Then, the LN gateway and bridge LN node should wait for the funding transaction to reach a specified number of confirmations on the blockchain (generally 3 confirmations). After reaching the specified depth, the LN gateway and the bridge LN node exchange \textit{funding\_locked} messages which finalizes the channel opening (\#12-13 in Fig. \ref{fig:openchannel}).

\end{itemize}

\subsection{Sending a Payment}
\label{sec:sendingpayment}

Similar to the channel opening, we incorporate the IoT device in threshold operations to authorize a payment sending. The same threshold schemes are utilized and LN's current payment sending protocol is modified. The steps of the proposed payment sending protocol is depicted in Fig. \ref{fig:sendpayment} and elaborated below:

\begin{figure}[h]
    \centering
    % \vspace{-1mm}
    \includegraphics[width=\linewidth]{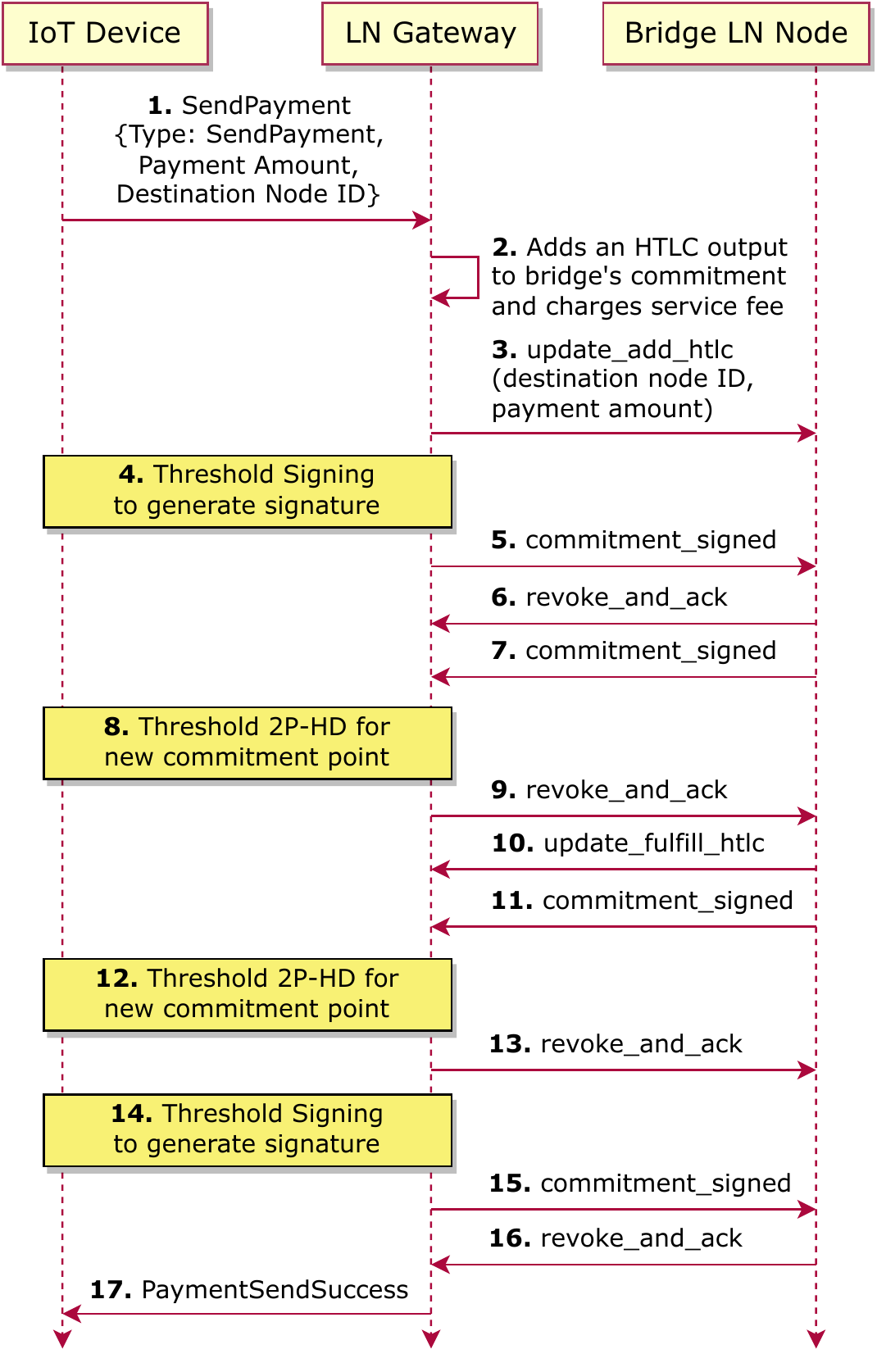}
    \vspace{-9mm}
    \caption{Protocol steps for sending a payment.}
    \label{fig:sendpayment}
    \vspace{-2mm}
\end{figure}

\begin{itemize}[leftmargin=*]

    \item \textbf{Payment Sending Initiation}: To request a payment sending, IoT device sends a \textit{SendPayment} message (\#1 in Fig. \ref{fig:sendpayment}) to the LN gateway. This message has the following fields: \textit{Type: SendPayment, Payment Amount, Destination Node ID}. Here, we assume that the \textit{Destination Node ID} is either interactively provided to the IoT device in some form (i.e., QR code) by the vendor (i.e., toll gate) just before the payment or it is known by the IoT device in advance. 

    \item \textbf{Payment Processing at the LN Gateway}: 
    Upon receiving the request, the LN gateway adds an HTLC output to bridge LN node's version of the commitment transaction (\#2 in Fig. \ref{fig:sendpayment}). When preparing the HTLC, the LN gateway \textit{deducts a certain amount of fee} from the real payment amount the IoT device wants to send to the destination. Therefore, the remaining Bitcoin is sent with the HTLC. This fee is taken to incentivize the LN gateway to continue serving the IoT devices. The LN gateway then sends an \textit{update\_add\_htlc} message (\#3 in Fig. \ref{fig:sendpayment}) to actually offer to the HTLC to the destination LN node 
    which is first received by the bridge LN node and then other nodes on the payment path (destination LN node is not shown in Fig. \ref{fig:sendpayment} for simplicity). Here, the \textit{destination node ID}, specified by the IoT device, is embedded into the \textit{onion routing packet} which is sent with the \textit{update\_add\_htlc} message.

    \item \textbf{1st Commitment Round}: In LN, HTLCs always require two rounds of \textit{commitment\_signed} and \textit{revoke\_and\_ack}. First round is for invalidating the old channel state right before the HTLC is attached to the channel. The second one is to fulfill the HTLC to remove it from the channel. Thus, at this point, the LN gateway will initiate the first round by sending a \textit{commitment\_signed} message to the bridge LN node. For that, it jointly generates a signature with the IoT device in a (2,2)-threshold signing (\#4 in Fig. \ref{fig:sendpayment}). The HTLC signature is also jointly generated at this step. Once the commitment and HTLC signatures are generated, they are sent to the bridge LN node in the \textit{commitment\_signed} message (\#5 in Fig. \ref{fig:sendpayment}). These signatures will enable bridge LN node to spend the new commitment transaction and the HTLC output. The bridge LN node responds to this message by first sending a \textit{revoke\_and\_ack} then a \textit{commitment\_signed} message (\#6-7 in Fig. \ref{fig:sendpayment}). Symmetrically, now the LN gateway will send a \textit{revoke\_and\_ack} message but before that we propose the LN gateway and the IoT device to  thresholdize the new commitment point in a 2P-HD (\#8 in Fig. \ref{fig:sendpayment}). Once this is done, the LN gateway sends the \textit{revoke\_and\_ack} message (\#9 in Fig. \ref{fig:sendpayment}).

    \item \textbf{Fulfilling the HTLC}: Now, the next step is for bridge LN node to fulfill the HTLC with a similar symmetric \textit{commitment\_signed} \& \textit{revoke\_and\_ack} round. Thus, it sends an \textit{update\_fulfill\_htlc} message to the LN gateway (\#10 in Fig. \ref{fig:sendpayment}) then a \textit{commitment\_signed} message (\#11 in Fig. \ref{fig:sendpayment}) to initiate the second round of commitments. This will be followed by the LN gateway sending a \textit{revoke\_and\_ack} message. However, the LN gateway first performs a 2P-HD with the IoT device to thresholdize the new commitment point (\#12 in Fig. \ref{fig:sendpayment}). After sending the \textit{revoke\_and\_ack} message (\#13 in Fig. \ref{fig:sendpayment}), the next step is to send the \textit{commitment\_signed} message. To generate the signature, the LN gateway performs a (2,2)-threshold signing (\#14 in Fig. \ref{fig:sendpayment}) with the IoT device then sends the \textit{commitment\_signed} message (\#15 in Fig. \ref{fig:sendpayment}). Finally, the bridge LN node replies with a \textit{revoke\_and\_ack} message to irrevocably fulfill the HTLC (\#16 in Fig. \ref{fig:sendpayment}).

    \item \textbf{Notifying IoT}: Now that the payment is successfully sent, the LN gateway can notify the IoT device of the successful payment by sending a \textit{PaymentSendSuccess} message (\#17 in Fig. \ref{fig:sendpayment}).

\end{itemize}

\subsection{Receiving a Payment}
\label{sec:receivingpayment}

Receiving payments on the channel is a bit different than other channel operations as it does not require the IoT device to send a request to the LN gateway. Rather, an LN node on the Internet initiates a payment to the IoT device which needs to be received on the channel that the LN gateway opened for it. Normally, when sending a payment on LN, it is enough to only specify the recipient's LN public key. If the recipient has multiple channels that can receive the payment, then the payment might end up at any of them. The exact channel that will receive the payment is internally decided by LN's routing algorithm. For our context, since the LN gateway has multiple channels each of which might be serving different IoT devices, receiving a payment on a specific channel becomes an important problem to tackle. In this direction, a sender LN node can force its payment to take a predetermined path which is also known as the \textit{source routing}. By querying the possible routes to the recipient node, a sender LN node can decide all the channels to use before sending the payment. We propose that the LN gateway \textit{does not charge a service fee} for receiving payments. The steps of the proposed protocol is shown in Fig. \ref{fig:receivepayment} and explained in detail below:

\begin{figure}[h]
    \centering
    \vspace{-1mm}
    \includegraphics[width=\linewidth]{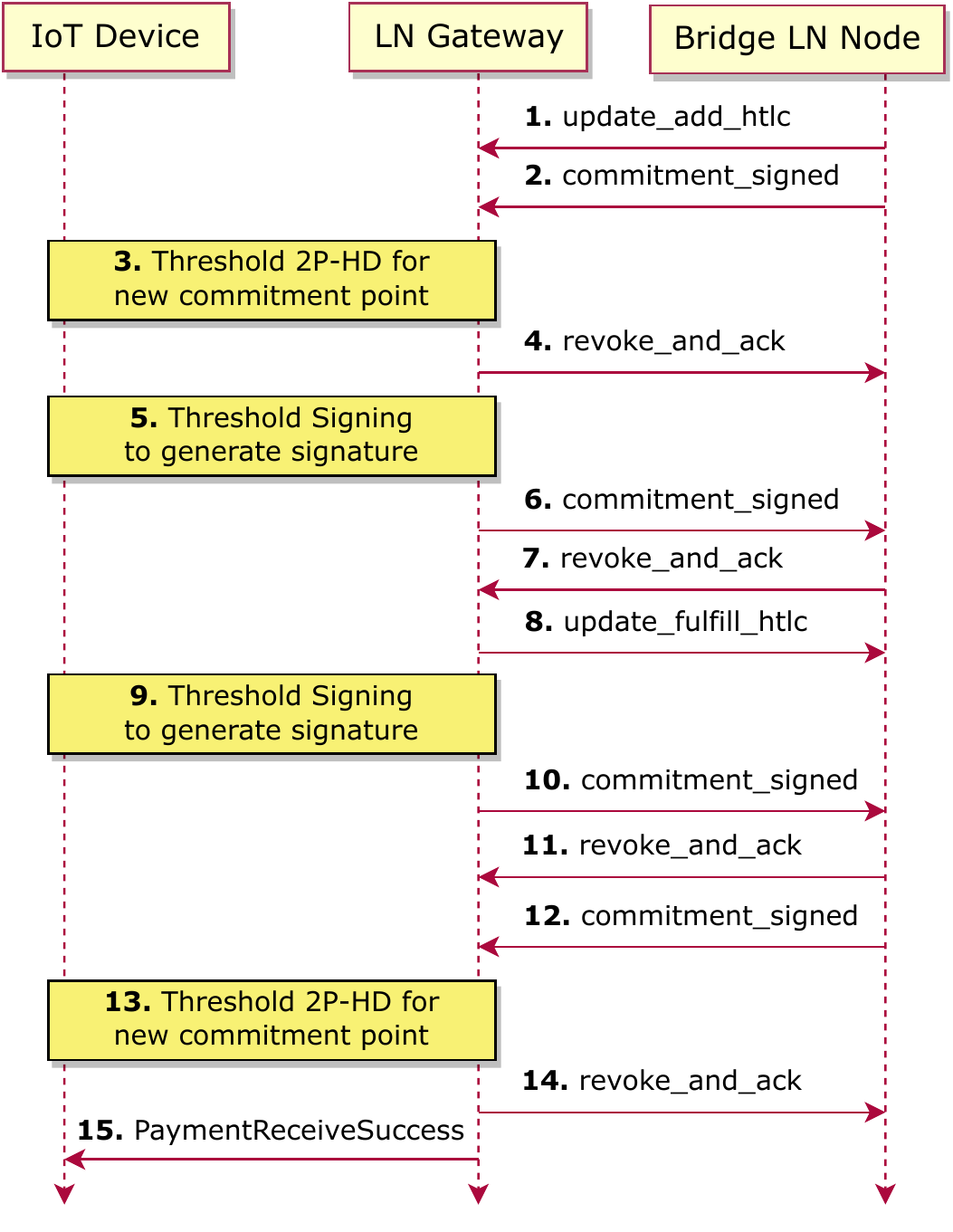}
    \vspace{-9mm}
    \caption{Protocol steps for receiving a payment.}
    \label{fig:receivepayment}
    \vspace{-1mm}
\end{figure}

\begin{itemize}[leftmargin=*]

    \item \textbf{Calculating a Route:} We assume that the sender LN node already knows to which IoT device to initiate the payment and the channel ID belonging to that IoT device (i.e., the channel the LN gateway opened to the bridge LN node for this specific IoT device). Knowing the channel ID, the sender LN node can calculate a route from its LN node to the LN gateway's LN node where the channel belonging to the IoT device is the last channel on the route. After calculating the route, the sender LN node can prepare a \textit{key send} payment that will use this specific route.
    
    \item \textbf{1st Commitment Round:} The sender LN node can initiate the payment by sending an \textit{update\_add\_htlc} message (\#1 in Fig. \ref{fig:receivepayment}) to the LN gateway which will be relayed to all the nodes on the payment path including the bridge LN node (sender LN node is not shown in Fig. \ref{fig:receivepayment} for simplicity). At this stage, similar to the payment sending case, there are going to be two rounds of \textit{commitment\_signed} and \textit{revoke\_and\_ack}. The bridge LN node initiates the first round by sending a \textit{commitment\_signed} message to the LN gateway to commit the initial changes on the channel (\#2 in Fig. \ref{fig:receivepayment}). Here, before sending a \textit{revoke\_and\_ack} message, the LN gateway performs a threshold 2P-HD (\#3 in Fig. \ref{fig:receivepayment}) with the IoT device to thresholdize the next commitment point. Then, it sends the \textit{revoke\_and\_ack} message (\#4 in Fig. \ref{fig:receivepayment}). Now, the LN gateway needs to send a \textit{commitment\_signed} message to the bridge LN node. Thus, to generate the signature, it performs a (2,2)-threshold signing with the IoT device (\#5 in Fig. \ref{fig:receivepayment}). Then, the \textit{commitment\_signed} message is sent (\#6 in Fig. \ref{fig:receivepayment}) and a \textit{revoke\_and\_ack} message is received (\#7 in Fig. \ref{fig:receivepayment}) from the bridge LN node.
    
    \item \textbf{Fulfilling the HTLC:} Messages \#2-7 made sure that the old channel state is invalidated so the bridge LN node cannot pretend the HTLC never existed. Now, the next step is to fulfill the HTLC with a similar symmetric \textit{commitment\_signed} \& \textit{revoke\_and\_ack} round. Thus, the LN gateway first sends an \textit{update\_fulfill\_htlc} message (\#8 in Fig. \ref{fig:receivepayment}). Then, the LN gateway and the IoT device generate a signature in a (2,2)-threshold signing (\#9 in Fig. \ref{fig:receivepayment}) which is sent to the bridge LN node in a \textit{commitment\_signed} message (\#10 in Fig. \ref{fig:receivepayment}). Bridge LN node responds by first sending a \textit{revoke\_and\_ack} message (\#11 in Fig. \ref{fig:receivepayment}) then a \textit{commitment\_signed} message (\#12 in Fig. \ref{fig:receivepayment}). Finally, before sending the \textit{revoke\_and\_ack} message, the LN gateway again performs a threshold 2P-HD with the IoT device to thresholdize the next commitment point (\#13 in Fig. \ref{fig:receivepayment}). It then sends the \textit{revoke\_and\_ack} message (\#14 in Fig. \ref{fig:receivepayment}) which fulfills the HTLC irrevocably.
    
    \item \textbf{Notifying IoT:} Now, it is a good time for the LN gateway to let the IoT device know that the payment is successfully received. Thus, it sends a \textit{PaymentReceiveSuccess} message to the IoT device.

\end{itemize}

\subsection{Channel Closing Process}
\label{sec:channelclosing}

A channel in LN is closed either unilaterally by one of the parties broadcasting its most recent commitment transaction to the blockchain or closed mutually by both parties agreeing on the closing fee. In our case, all 3 parties of the channel namely; the IoT device, the LN gateway, and the bridge LN node can close the channel. We explain all three cases separately below:

\subsubsection{IoT Device Channel Closure}

When the IoT device would like to close the channel, it follows the proposed protocol below.

\begin{itemize}[leftmargin=*]

\item \textbf{IoT Device Channel Closing Request:} The IoT device sends a \textit{ChannelClosingRequest} message to the LN gateway. 

\item \textbf{Mutual Close:} The LN gateway has two options to close the channel which are \textit{unilateral} or \textit{mutual} close. In mutual close, the LN gateway and the bridge LN node first exchange \textit{shutdown} messages and then start negotiating on the channel closing fee. For this, they start exchanging \textit{closing\_signed} messages. This message includes the offered fee and offering party's signature. Thus, each time the LN gateway sends a \textit{closing\_signed} message, it has to perform a (2,2)-threshold signing with the IoT device to generate the signature. Once the closing fee is agreed upon, the closing transaction is broadcast to the blockchain by the LN gateway. 

\item \textbf{Unilateral Close:} In the unilateral close case, the LN gateway just broadcasts its most recent commitment transaction to the Bitcoin network after getting it (2,2)-threshold signed with the IoT device. Once the broadcast transaction is mined, the channel is closed and everyone's funds in the channel settle in their respective Bitcoin addresses.

\end{itemize}

We propose the on-chain fee for both cases to be paid by the IoT device since the channel closing was requested by the IoT device. The fee is deducted from the IoT device's Bitcoin address.

\subsubsection{LN Gateway Channel Closure}

The LN gateway can also initiate the channel closing if it wants to close IoT device's channel for any reason. The steps are very similar to IoT device channel closure case and explained below:

\begin{itemize}[leftmargin=*]

    \item \textbf{LN Gateway Channel Closing Request:} The LN gateway sends a \textit{ChannelClosingRequest} message to the IoT device to show its intention to close the channel.
    
    \item \textbf{Closing the Channel:} The LN gateway can close the channel unilaterally or mutually. For either case, it needs the IoT device to participate in a (2,2)-threshold signing. The steps for closing the channel are exactly the same of the IoT device channel closure case explained above. The only difference is that, since now the channel closing is requested by the LN gateway, the on-chain fee is paid by the LN gateway by deducting the fee from its Bitcoin address. 
    
\end{itemize}

\subsubsection{Bridge LN Node Channel Closure}

The bridge LN node can close the channel unilaterally or mutually. A mutual close from the bridge LN node will trigger a fee negotiation phase with the LN gateway which involves exchanging \textit{closing\_signed} messages. Since this requires the IoT device to be online and participate in (2,2)-threshold signing with the LN gateway, the bridge LN node might have to close the channel unilaterally if the IoT device is not online at the time of the mutual close attempt. For the unilateral close, the bridge LN node can just broadcast its most recent commitment transaction to the blockchain.

\begin{figure*}[t]
    \centering
    % \vspace{-2mm}
    \includegraphics[width=0.8\linewidth]{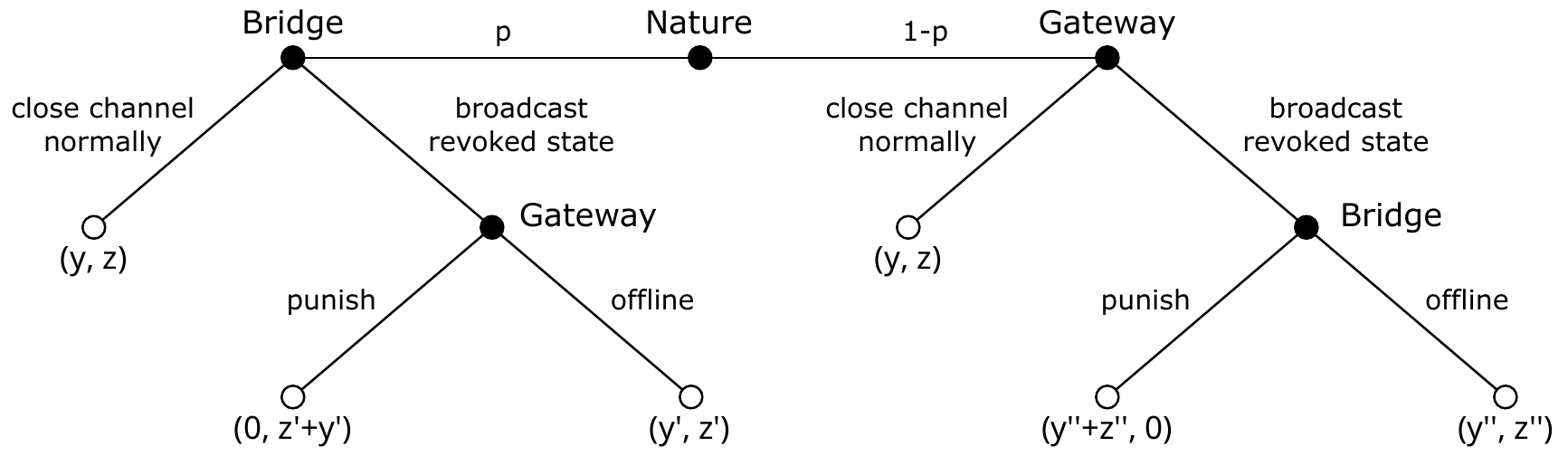}
    \vspace{-4mm}
    \caption{Extensive form of the behavioral game for channel closures between the LN gateway and the bridge LN node.}
    \vspace{-4mm}
    \label{fig:game_theory_nocollusion}
\end{figure*}

\subsubsection{Behavioral Analysis of Channel Closures}
\label{sec:behavioral}

In our preliminary version of this work \cite{kurt2021lngate}, the protocol was only capable of unidirectional payments. With unidirectional payments, the LN gateway and the bridge LN node cannot make profit by broadcasting revoked states or colluding with each other. However, enabling bidirectional payments where the IoT device can also receive payments from other nodes in LN requires revisiting the previous analysis. The problem is, if the IoT device receives a payment, it will have some old states in which it has less money. Thus, the LN gateway or the bridge LN node can publish these old channel states to the blockchain to cheat and profit. To better illustrate these cases and have a formal analysis, we will use a \textit{sequential game} approach using \textit{game theory} and present the actions of each party that leads to various outcomes. Then, we will present the \textit{Nash equilibrium} for each case using \textit{backward induction}. We assume that all players act rationally to maximize their profits.

\vspace{1mm}
\noindent \textbf{Game Model}: The game theoretical model in an extensive form, i.e., tree-form, is shown in Fig. \ref{fig:game_theory_nocollusion} and set up as follows: There are two players in this game: The LN gateway (Gateway) and the bridge LN node (Bridge). Both the Gateway and Bridge can start the game.

With probability \textit{p}, Bridge starts the game. So, at its first decision node, Bridge has to make a decision to close the channel normally or broadcast a revoked state to the blockchain. If Bridge plays to close the channel normally, the game ends at this terminal node with the payoffs of $y$ BTC and $z$ BTC for Bridge and Gateway, respectively (i.e., ($y$, $z$) BTC). Furthermore, we assume that the IoT device (IoT) will have $x$ BTC for this case. Here, we can write the following equation:

\begin{equation}
    x + y + z = C
\end{equation}

\noindent where $C$ is the channel balance.

\begin{figure*}[t]
    \centering
    % \vspace{-2mm}
    \includegraphics[width=\linewidth]{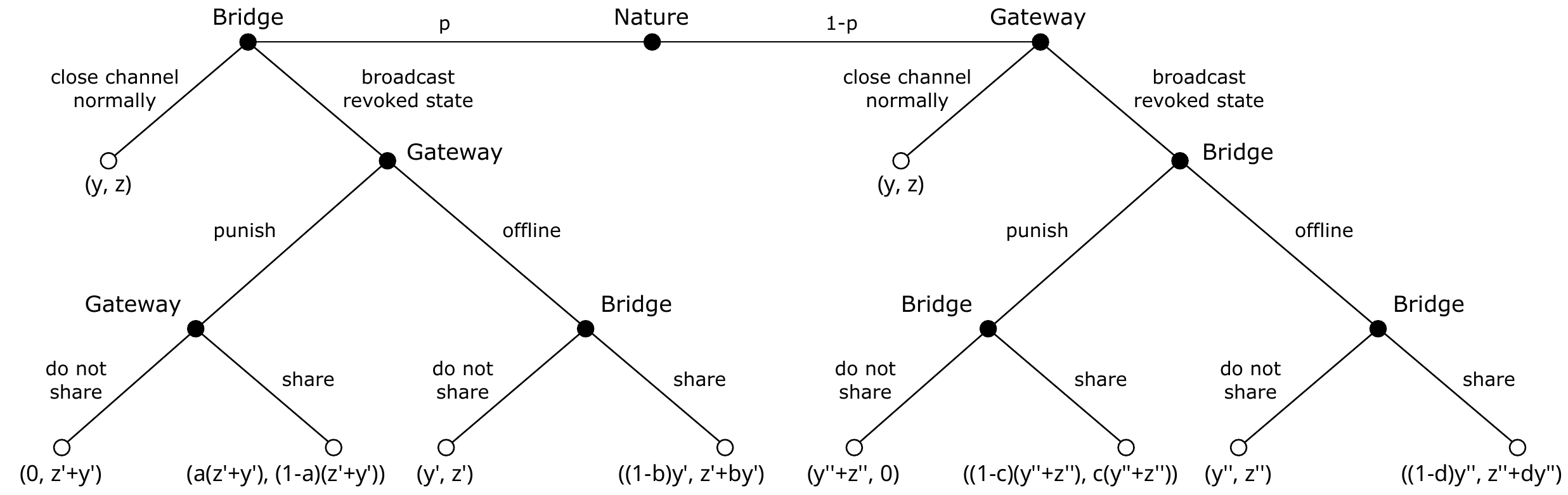}
    \vspace{-8mm}
    \caption{Extensive form of the collusion game between the LN gateway and the bridge LN node.}
    \vspace{-4mm}
    \label{fig:game_theory_collusion}
\end{figure*}

If Bridge chooses to broadcast a revoked state, the Gateway gets to play at its first decision node in this game. At this node, normally Gateway will punish the Bridge as that is the way the LN protocol works. As explained before, if an LN node broadcasts an old state, its funds will be automatically swept by the counterparty. However, since we are examining the behavior where Gateway and Bridge are trying to steal funds from the IoT, there is a chance that Gateway will go offline before Bridge broadcasts the old state. Or, it might also happen that Gateway goes offline due to a power/Internet outage which Bridge can see and try to exploit. Thus, at this node, Gateway can either punish the Bridge or be offline.

If Gateway plays to punish the Bridge, Bridge's funds in this old state will be taken by the Gateway as proposed in our protocol description in Section \ref{sec:LN_modified}. If the channel balances of IoT, Bridge and Gateway are ($x'$, $y'$, $z'$) BTC respectively in this old state, Gateway will have $z'+y'$ BTC after punishing the Bridge. Here, we can write the following equations:

% \vspace{1mm}
\begin{equation}
\begin{aligned}
   & x' + y' + z' = C \\
   & y' > y \\
   & z' \leq z \\  
   & x' < x
\end{aligned}
\end{equation}
% \vspace{1mm}

\noindent because we know that Bridge has more funds in the broadcast revoked state and Gateway either has the same amount of funds or less. Gateway cannot have more funds in an old state because its balance only increases when IoT sends payments. Gateway does not charge fees when IoT receives payments on the channel. Hence, the game ends at this terminal node with the payoffs ($0$, $z'+y'$) BTC. 

If Gateway plays to go offline instead, Bridge will end up with some extra funds. Thus, the end game payoffs will be ($y'$, $z'$) BTC. 

Game can also be started by the Gateway with probability $1-p$. Normally, Gateway does not have an incentive to broadcast a revoked state because it does not have any old states in which it has more funds. But, we still analyze it for the sake of completeness. In its decision node, Gateway decides to whether close the channel normally or broadcast an old state. If Gateway chooses to close the channel normally, the game ends at this terminal node with the same payoffs as before; ($y$, $z$) BTC. If Gateway chooses to broadcast an old state where the channel balances are ($x''$, $y''$, $z''$) BTC for IoT, Bridge and Gateway respectively, Bridge will have the options of whether to punish the Gateway or be offline. Here, similar as before, we can write the following equation:

\begin{equation}
    x'' + y'' + z'' = C
\end{equation}

If Bridge plays to punish the Gateway, Gateway's funds in this old state will be taken by Bridge as proposed in our protocol description in Section \ref{sec:LN_modified}. Thus, Bridge will have $y''+z''$ BTC after punishing the Gateway. Here, we can write the following equations:

% \vspace{1mm}
\begin{equation}
\begin{aligned}
   & y'' > y \\
   & z'' \leq z \\  
   & x'' < x
\end{aligned}
\end{equation}

One should note that, in this node, Gateway will get zero payoff. Thus, the game ends at this terminal node with the payoffs ($y''+z''$, $0$) BTC. 

If Bridge plays to be offline instead; unlike the previous scenario, Gateway will not end up with some extra funds. Instead, Bridge ends up with some extra funds. In that case, the end game payoffs will be ($y''$, $z''$) BTC. 

\vspace{1mm}
\noindent \textbf{Equilibrium Analysis}:

\newtheorem{theorem}{Theorem}
\begin{theorem}
In a simple behavioral game for channel closures where the Bridge plays first, the Nash equilibrium is \textit{“close channel normally”} for the Bridge, and \textit{“punish”} for the Gateway. Symmetrically, when the Gateway plays first, the Nash equilibrium is \textit{“close channel normally”} for the Gateway, and \textit{“punish”} for the Bridge.
\end{theorem}

\begin{proof}
For the proof of Bridge starting the game first, we apply the backward induction. Thus, we observe that Gateway plays to punish the Bridge at the terminal node since $z'+y' > z'$ from Equation (2). Similarly, Bridge obtains a better payoff if it plays to close the channel normally because $y>0$. Hence, it implies that the optimal solution for this game is that Bridge chooses \textit{“close channel normally”} and Gateway chooses \textit{“punish”}. That is the Bridge chooses to close the channel normally. In the event that Bridge chooses to broadcast a revoked state, Gateway plays to punish the Bridge. This equilibrium implies that the Bridge will always close the channel normally which is the intended behavior in LN.

In a similar fashion, when the Gateway starts the game first, we observe that Bridge plays to punish the Gateway as $y''+z'' > y''$ from Equation (4) using backward induction. As expected, the Gateway becomes better off by playing to close the channel normally since $z>0$. Therefore, the optimal solution for this game is that Gateway chooses \textit{“close channel normally”} and Bridge chooses \textit{“punish”}. Again, if the Gateway chooses to broadcast a revoked state, Bridge plays to punish the Gateway. Finally, we obtain that the Gateway will always close the channel normally which is again the intended behavior in LN. 
\end{proof}

\section{Security Analysis}
\label{sec:security}
In this section, we show how our proposed protocol addresses the attacks mentioned in Section \ref{sec:threatmodel}.

\subsection{Collusion Attacks}
\label{sec:gametheory}

Broadcasting revoked states can be exploited by the Gateway and Bridge when they collude with each other to increase their chances of stealing money from the IoT and share that profit when they are successful. It is important to note that when the Gateway and Bridge are controlled by the same entity, this can cause IoT to lose funds. However, as explained in Section \ref{sec:channelopening}, the Gateway and Bridge are not controlled by the same entity since the Bridge is chosen by the IoT per our approach. Building on the analysis in Section \ref{sec:behavioral}, we formally analyze the collusion game as follows:

\begin{theorem}
In a collusion game where the Bridge plays first, the Nash equilibrium is \textit{“close channel normally”} for the Bridge and \textit{“punish”}, \textit{“do not share”} for the Gateway. When the Gateway plays first, the Nash equilibrium is \textit{“close channel normally”} for the Gateway and \textit{“punish”}, \textit{“do not share”} for the Bridge.
\end{theorem}

\begin{proof}
Similar to the proof of Theorem 1, we use the backward induction. The only difference from the previous proof is that, we include an additional subgame, i.e., the end subgame, in Fig. \ref{fig:game_theory_collusion} for both cases. For the case of Bridge starting the game first, we apply the backward induction and check Gateway's and Bridge's payoffs at this end subgame (their second decision nodes). At its second decision node, Gateway chooses not to share profits since $z'+y' > (1-a)(z'+y')$. Furthermore, Bridge also chooses not to share profits at its second decision node since $y' > (1-b)y'$. Here, \textit{a} and \textit{b} represent the ratios used to share the profits. After that, we continue the backward induction procedure, we observe that Gateway plays to punish the Bridge as $z'+y' > z'$ and Bridge chooses to close the channel normally because $y>0$ as given in the proof of Theorem 1. Hence, the optimal solution for this game is that the Bridge chooses \textit{“close channel normally”}, the Gateway chooses \textit{“punish”} and \textit{“do not share”}. In other words, the Bridge chooses to close the channel normally. In the even that Bridge chooses to broadcast a revoked state, the Gateway plays to punish the Bridge and not share the profits. This equilibrium implies that the Bridge will not broadcast a revoked state as it results in Gateway getting all its funds and not share anything back. Thus, the Bridge will always close the channel normally, i.e., our protocol is secure against the collusion attacks.

For the case of the Gateway starting the game first, one should again apply the backward induction and check Bridge's payoffs at both end subgames. For both punish and offline cases, Bridge chooses not to share the profits since $y''+z'' > (1-c)(y''+z'')$ and $y'' > (1-d)y''$. Similar as before, \textit{c} and \textit{d} represent the ratios used to share the profits. Next, we observe that Bridge plays to punish the Gateway as $y''+z'' > y''$ from Equation (4) and Gateway chooses to close the channel normally because $z>0$ which was also shown with the proof of Theorem 1. Similar to the first case, we then find that the optimal solution for this game is that the Gateway chooses \textit{“close channel normally”}, the Bridge chooses \textit{“punish”} and \textit{“do not share”}. This implies that the Gateway chooses to close the channel normally. In the event that the Gateway chooses to broadcast a revoked state, the Bridge plays to punish the Gateway and not share the profits. The same reasoning about the equilibrium applies here and hence the Gateway will always close the channel normally, which again proves that our protocol is secure against the collusion attacks.
\end{proof}

\subsection{Stealing IoT Device's Funds}
Using (2,2)-threshold signatures for the LN operations secure the IoT device's funds in the channel since the LN gateway: 1) cannot send IoT device's funds in the channel to other LN nodes without generating proper HTLC and commitment signatures with the IoT device in a (2,2)-threshold signing; 2) cannot cause loss of IoT device's funds by broadcasting revoked states as shown in Section \ref{sec:gametheory} and; 3) cannot cause loss of IoT device's funds by colluding with the bridge LN node again as shown in Section \ref{sec:gametheory} with the game theoretic security analysis. If we used LN's original signing mechanism, the LN gateway could move IoT device's funds in the channel without needing a signature from the IoT device. Consequently, the usage of (2,2)-threshold schemes along with the proposed modifications to LN's commitment transactions prevent IoT device from losing any funds.

\subsection{Ransom Attacks}
This is an attack where the LN gateway deviates from the protocol description. To put it with some examples, the LN gateway can say to the IoT device: ``I will perform your channel closing request only if you pay me X amount of Bitcoins" or; ``from now on, I will execute your payment sending requests only if you accept to pay an 10\% increased service fee". This essentially turns into a game where the IoT device's best move is to reject the ransom attempt and just wait. Then, the LN gateway would just hold the IoT device's funds hostage for as long as it can in an attempt to deter the IoT device. This is a \textit{deadlock} case where both parties just wait. It is clear that a \textit{rational} LN gateway has no incentive to perform ransom attacks as it does benefit from these attacks assuming that the IoT device acts rationale. The best course of action for the LN gateway is to continue serving the IoT device and keep collecting the service fees. Thus, our proposed protocol protects the IoT device against ransom attacks.

\begin{table*}[h]
  \begin{center}
%   \vspace{-2mm}
    \caption{Execution Times of Channel Opening and Closing for ``WiFi", ``BLE" and ``No IoT" Cases}
     \vspace{-4mm}
    \label{tab:communication1}
    \resizebox{0.84\linewidth}{!}{
    \begin{tabular}{|c|c|c|c|c|c|}
    \hline
    & \textbf{from} & \textbf{to} & \textbf{WiFi Delay} & \textbf{BLE Delay} & \textbf{No IoT Delay} \\
   \hline
   \textbf{Channel Opening}  & LNGate$^2$ & VIOLETWALK &  1.73 s  &  2.85 s  &  0.32 s  \\ \hline
   \textbf{Channel Closing}  & LNGate$^2$ & VIOLETWALK  &  0.91 s &  1.35 s  &  0.18 s  \\ \hline
    \end{tabular}
    }
   \vspace{-5mm}
  \end{center}
\end{table*}

\begin{table*}[h]
  \begin{center}
%   \vspace{-2mm}
    \caption{Execution Times of Payment Operations for ``WiFi", ``BLE" and ``No IoT" Cases}
     \vspace{-6mm}
    \label{tab:communication2}
    \resizebox{\linewidth}{!}{
    \begin{tabular}{|c|c|c|c|c|c|}
    \hline
    & \textbf{from} & \textbf{to} & \textbf{WiFi Delay} & \textbf{BLE Delay} & \textbf{No IoT Delay} \\
   \hline
  \textbf{Sending Invoice Payment}      & LNGate$^2$ & VIOLETWALK &  1.07 s  &  2.75 s  &  0.31 s  \\ \hline
   \textbf{Sending Key Send Payment}    & LNGate$^2$ & VIOLETWALK &  1.01 s  &  2.97 s  &  0.33 s  \\ \hline
   \textbf{Receiving Key Send Payment}  &  VIOLETWALK & LNGate$^2$  &  1.08 s  &  2.93 s  &  0.3 s  \\ \hline
    \end{tabular}
    }
   \vspace{-5mm}
  \end{center}
\end{table*}

\section{Evaluation}
\label{sec:evaluation}
This section describes the experiment setup to implement the proposed approach and presents the performance results.

\subsection{Experiment Setup and Metrics}

To evaluate the proposed protocol, we implemented it by modifying the source code of Core Lightning v0.9.3 \cite{clightning} which is one of the implementations of the LN protocol written in C. The \textit{hsmd} and \textit{bitcoin} modules of Core Lightning were modified. The hsmd module manages the cryptographic operations and controls the funds in the channel. As the name suggests, the bitcoin module handles the Bitcoin script, signature and transaction routines. Specifically, we thresholdized the \textit{funding\_pubkey} and the associated Bitcoin private key as explained in Section \ref{sec:channelopening}. To the best of our knowledge, this is the first-ever work that implemented threshold cryptography for LN. Our implementation is publicly available in our GitHub repository at \url{https://github.com/startimeahmet/lightning}.

In the experiments, we used WiFi (IEEE 802.11n) and Bluetooth Low Energy (BLE) as the main communication protocol between the IoT device and the LN gateway. The purpose of performing the experiments with both WiFi and BLE is to investigate how different wireless technologies perform with our approach. Thus, it is essential for the rest of the components of the experiment setup to stay the same while changing the wireless communication method. In this direction, we created two experiment setups that we will call \textit{WiFi setup} and \textit{BLE setup}. For all experiments, we used Bitcoin's Testnet as the base-layer. Similar to Bitcoin's Mainnet, Testnet network consists of real Bitcoin nodes all around the world. However, Testnet Bitcoins do not have a monetary value unlike Mainnet which makes it more suitable for development and testing purposes.

Both setups use Raspberry Pi 4 Model B as the IoT device which is equipped with on-board dual-band IEEE 802.11b/g/n/ac wireless, Bluetooth 5.0 and BLE. For the LN gateway, we used a desktop computer with 2 Intel(R) Xeon(R) E5-2690 v4 CPUs and 32 GB of RAM. The desktop computer was connected to the Internet through a Gigabit Ethernet connection inside Florida International University campus. It ran our modified version of the Core Lightning for the full LN node. For the full Bitcoin node, it ran \textit{bitcoind} \cite{bitcoind} which is one of the most widely used implementations of the Bitcoin protocol. For the threshold operations, we used \textit{Gotham city} \cite{gothamcity} which is a decentralized client/server Bitcoin wallet application that utilizes 2-party threshold ECDSA. Gotham city consists of \textit{Gotham client} and \textit{Gotham server}. Gotham client is the wallet application and was run on the desktop machine. Gotham server is a RESTful web service acting as a server for the threshold operations and was run on the Raspberry Pi. The other way is also possible (server on the desktop machine, client on the Pi) but it was easier to run the server on the Pi in our setup. Additionally, we created another LN node running the original Core Lightning v0.9.3 software in a DigitalOcean droplet in New Jersey to use in our experiments.

For the WiFi setup, we connected a TP-Link TL-WN722N 150Mpbs Wireless USB Adapter to the desktop machine. Using this adapter, we created a WiFi hotspot which the Raspberry Pi connected to. In this way, the Pi directly communicated with the desktop machine over the WiFi (IEEE 802.11n) connection.

For the BLE setup, we connected an Asus USB-BT500 Bluetooth 5.0 USB Adapter (supports BLE) to the desktop machine. Now, even though both the desktop machine and the Raspberry Pi have BLE connectivity, they cannot perform the threshold operations over BLE because the Gotham city is designed to work over TCP/IP. To overcome this issue, we enabled TCP/IP to work over BLE on the devices which is known as IP over BLE. A bunch of configuration had to be done for this to work such as installing \textit{bluez} package, creating a Personal Area Network, configuring the master node and the slave node and more. The further details of our experiment setups, additional technical details and the tools \& scripts we used are provided in our GitHub page at \url{https://github.com/startimeahmet/LNGate2}.

To assess the performance of our protocol, we used the following metrics: 1) \textit{Time} which refers to the communication and computational delays of the proposed protocol; 2) \textit{Cost} which refers to the monetary costs associated with our proposed protocol; 3) \textit{Bandwidth} which refers to the network usage of the IoT device and the minimum required bandwidth (data rate) for the IoT device for timely LN operations; 4) \textit{Scalability} which refers to the scalability of the proposed protocol for increasing number of IoT devices and payments; 5) \textit{Energy} which refers to the energy consumption of the IoT device when using the proposed protocol.

To compare our approach to a baseline, we considered the case where the LN gateway performs the LN operations by itself such as sending and receiving a payment. In other words, no IoT device is present, and all LN tasks are solely performed by the LN gateway. We will refer to it as \textit{no IoT case} in the next sections.

\begin{table}[h]
    \centering
    \vspace{-3mm}
    \caption{Pure Computation Times}
    \vspace{-3mm}
    \label{tab:computational}
    \resizebox{\linewidth}{!}{
    \begin{tabular}{|c|c|c|c|c|c|}
        \hline
        \multirow{2}{*}{\textbf{\makecell{AES \\ Encryption}}} & \multirow{2}{*}{\textbf{\makecell{HMAC \\ Calculation}}} & \multicolumn{2}{c|}{\textbf{\makecell{(2,2)-Threshold \\ Key Generation}}} & \multicolumn{2}{c|}{\textbf{\makecell{(2,2)-Threshold \\ Signing}}} \\
        \cline{3-6}
        & & \textbf{IoT} & \textbf{Desktop} & \textbf{IoT} & \textbf{Desktop} \\
        \hline
        15 ms & $<$ 1 ms & 1.78 s & 0.53 s & 166 ms & 74 ms \\
        \hline
    \end{tabular}
    }
    \vspace{-6mm}
\end{table}

\subsection{Communication and Computational Delays}
\label{sec:overheadresults}

We first assessed the communication and computational delays of our proposed protocol. Computational overhead of running the protocol on the Raspberry Pi involves 3 different computations. These are the AES encryption of the protocol messages, HMAC calculations and (2,2)-threshold computations. We used Python's \textit{pycrypto} library to encrypt the protocol messages with AES-256 encryption. The encrypted data size for the messages was 24 bytes. For the authentication of the messages, we used HMAC. To calculate the HMACs, \textit{hmac module} in Python was used. Measuring the pure computation times of the (2,2)-threshold key generation and signing is a little bit tricky because Gotham client and Gotham server are run on separate devices. We can instead run both the client and the server on the same device and use localhost for the server to eliminate any real network traffic. In this direction, we ran the client and the server on the Pi as well as the desktop PC to measure the best and worst cases of the pure computation times. We present these measured computation times in Table \ref{tab:computational}. All values are an average of 30 runs of the respective operation.

As can be seen from the results, the overhead of the AES encryption and HMAC calculation are negligible. (2,2)-threshold key generation takes 1.78 seconds on the Pi and 0.53 seconds on the desktop machine. Thus, the real delay is between these two values and it is not critical for LN operations since it is done 1-time at channel opening as explained in Section \ref{sec:channelopening}. The (2,2)-threshold signing on the other hand takes 166 ms on the Pi and 74 ms on the desktop machine which is much quicker than key generation.

We then measured the execution time of our protocol for 5 different LN operations using WiFi and BLE as the communication method between the Pi and the desktop machine. The 5 LN operations we considered are: 1) Channel opening, 2) Channel closing, 3) Sending an invoice payment, 4) Sending a key send payment, 5) Receiving a key send payment. Execution time of an LN operation can be broken down to the sole execution time of the LN operation at the LN gateway plus the execution time of the (2,2)-threshold operations between the IoT device and the LN gateway. The results are presented in Table \ref{tab:communication1} and \ref{tab:communication2} and they are used in other experiments to evaluate the timeliness of the protocol. Note that, the channel opening and channel closing delays in Table \ref{tab:communication1} do not include the confirmation times on the blockchain.

The \textit{`from'} and \textit{`to'} fields in the Table \ref{tab:communication1} and \ref{tab:communication2} are the LN node aliases. For example, the second row of Table \ref{tab:communication2} should be interpreted as follows: A key send payment was sent from the node with alias \textit{LNGate$^2$} to the node with alias \textit{VIOLETWALK}. \textit{LNGate$^2$} is our node with the threshold modifications running on the desktop machine that is located in Florida. \textit{VIOLETWALK} is our node running the original LN software on a DigitalOcean droplet in New Jersey. Each delay value in Table \ref{tab:communication1} and \ref{tab:communication2} is an average of 30 executions of the respective LN operation for statistical significance.

We first present the channel opening and closing delays in Table \ref{tab:communication1} as they are 1-time operations and their execution times are not critical since they should be done in advance before the payment services are used. Therefore, blockchain confirmation delay of these operations also will not have impact on the actual payment transactions. It is important to also note here that the users can use any of their existing channels to make payments which alleviates the need to open a specific channel for every transaction consequently the need to wait for channel openings. As can be seen in Table \ref{tab:communication1}, the gap is not significant compared to no IoT case. Specifically, WiFi case has around 1.4 seconds of extra delay compared to the no IoT case which mostly comes from the computation time of the threshold key generation. Channel opening delay of the BLE case is around 1 second longer compared to the WiFi case which is normal due to BLE's limited bandwidth. For the channel closing, WiFi case is only 0.7 seconds slower than the no IoT case which is again mostly due to the computation time of the threshold signing operations. A mutual channel close involves several rounds of threshold signing for peers to agree on the closing fee as explained in Section \ref{sec:channelclosing}. BLE is slower by approximately 0.4 seconds than the WiFi due to its lower bandwidth.

Next, we show the results of payment sending and receiving in Table \ref{tab:communication2} which are of more importance to the usability of the protocol. As can be seen, regardless of the payment operation, the WiFi case took around 1 second while no IoT case took 0.3 seconds and BLE case took close to 3 seconds. To interpret these results, we need to look at how many times threshold signing is performed in an LN payment. As shown in Section \ref{sec:sendingpayment} and \ref{sec:receivingpayment}, each HTLC requires two threshold signing. In our experiments, we realized that sometimes a single payment creates multiple HTLCs resulting in performing more than two threshold signing. Specifically, invoice and key send payments in this experiment had four threshold signing operations which took between (296, 664) ms. This tells us that the extra 0.7 seconds delay in the WiFi case mostly came from the computations of the threshold signing operations. The communication delays constitute a very small part of WiFi's overall delay.

We can also see that BLE adds roughly another 1.8 seconds to the delay compared to WiFi due to its limited data rates for transmitting threshold messages. Nevertheless, a delay of 2.97 seconds to send a key send payment under BLE can be considered fast enough for most scenarios. In particular, WiFi and BLE delays are comparable to or even less than that of any typical credit card payment delay which can take two to three seconds to get approved. This time increases further with two factor authentication services for security (e.g., more than 10 seconds) \cite{reese2019usability}.

\subsection{Cost Analysis}
We consider and analyze the following costs associated with our proposed protocol: 1) The on-chain fees for channel opening and closing; 2) The fees that are charged by the LN gateway for IoT device's payments; 3) Forwarding fees that are charged by the nodes on a payment route (i.e., bridge LN node).

According to our experiment results; a channel opening transaction signed with (2,2)-threshold ECDSA, incurs a fee of only 222 satoshi when it is desired for the transaction to be included in the next block\footnote{See this channel opening transaction which is signed using (2,2)-threshold ECDSA: \url{https://blockstream.info/testnet/tx/f2dfec159f66be6b785c97b247c44d6efd6fc9cd40a1b2d800386cec450f797a}}. Similarly, a mutual channel closing transaction signed with (2,2)-threshold ECDSA, costs only 183 satoshi\footnote{See this mutual channel closing transaction signed with (2,2)-threshold ECDSA: \url{https://blockstream.info/testnet/tx/9714d8796425b472ccfa8e049b83f72d59d1505cc805d9a3588fdc8b6865213d}}. At current Bitcoin price of \$30,000, these fees correspond to \textit{6.6 cents} and \textit{5.5 cents} respectively which are basically negligible considering these operations are 1-time.

LN gateway's service fee for payments entirely depends on the LN gateway's choice. Forwarding fees, again entirely depend on nodes' choices that are on a payment path. In LN, nodes charge a fixed fee each time they route a payment which is called the \textit{base fee} \cite{lnroutingfees}. There is also \textit{fee per satoshi} that the nodes charge proportional to the satoshi value of the payments they route \cite{lnroutingfees}. To measure an approximate value for the forwarding fee, we sent 30 LN payments using different routes for each. The average of the forwarding fees was \textit{2 satoshi} which is again negligible.

Most importantly; the channel opening \& closing fees and forwarding fees do not change with the introduction of (2,2)-threshold ECDSA. These values we measured are exactly the same for a regular LN node running the original LN software. Therefore, our protocol does not bring extra cost overhead for the LN operations.

\subsection{Network Usage and Bandwidth Analysis}
In this experiment, we investigate: 1) the \textit{network usage} of the IoT device; 2) the \textit{minimum} \textit{required bandwidth} for the IoT device to timely complete the LN operations with the LN gateway. Our motivation for finding the minimum required bandwidth is to understand if very low bandwidth IoT devices can participate in our protocol. To measure the network usage, we used Wireshark packet analyzer software. We captured the packets on a specific network interface, e.g., \texttt{wlxec086b180e1b} for WiFi, \texttt{pan0} for BLE. After the capture, using the \textit{endpoint statistics} of Wireshark, we collected the number of packets and bytes information. We considered the same 5 LN operations which are: 1) Channel opening, 2) Channel closing, 3) Sending an invoice payment, 4) Sending a key send payment, 5) Receiving a key send payment. All LN operations were performed 5 times and the average values were calculated.

The network usage of WiFi and BLE cases were the same and it ranged between ($\approx$ 16,000, $\approx$ 42,000) bytes where ($\approx$ 80, $\approx$ 200) network packets were exchanged between the IoT device and the LN gateway. Channel opening and closing were less intensive in terms of network usage compared to sending and receiving payments.

Now, the minimum required bandwidth for the IoT device can be calculated. If we focus on sending key send payments where the IoT device have to exchange around 42,000 bytes in 2.97 seconds (Table \ref{tab:communication2}), then the minimum required bandwidth is 113 kbps. This is a reasonable bandwidth requirement considering that even the low-power wireless technology 6LoWPAN supports a data rate of 250 kbps \cite{rfc6282}.

\subsection{Scalability Analysis}
To test the scalability of our approach, we performed additional experiments using increased number of Raspberry Pis instead of just 1. This setup consists of 5 Pis all associated with a separate LN node running on the desktop computer. Basically, the desktop machine ran 5 LN nodes and 5 Gotham clients to serve each Pi. On the other hand, each Pi ran a Gotham server to perform the LN operations with their corresponding LN nodes on the desktop machine.

To test the scalability of our approach, we conducted two different experiments: 1) Changed the number of sender IoT devices (i.e., Pis) where each of them is sending 100 payments at once to a single recipient LN node. The goal of this experiment is to investigate how increasing the number of Pis will impact the computational overhead on the LN gateway as it will now serve more Pis at the same time. In particular, we are interested to see how it will affect the overall delays; and 2) Changed the number of payments from a single sender IoT device to a single recipient LN node. The goal here is to see how increased number of payments impact the performance by keeping the computational overhead on the LN gateway constant by only serving one IoT device. For this experiment case, we also used an unmodified LN node as the sender node to create a baseline case. This way, we can compare the results with our approach to see if any change in the delays are due to LN's internal mechanisms or related to our approach. For each of these experiment cases, we created bash scripts to automate the process. Each script was run 30 times and average results were used for statistical significance. For the recipient LN node, we used our \textit{VIOLETWALK} node.

\begin{table}[h]
%  \vspace{-5mm}
  \begin{center}
  \vspace{-2mm}
    \caption{Effect of Increasing the Number of IoT Devices on Payment Delays}
    \vspace{-3mm}
    \label{tab:scalability1}
    \resizebox{0.95\linewidth}{!}{
    \begin{tabular}{|c|c|}
    \hline
       \textbf{Number of IoT Devices}  &  \textbf{Average Payment Delay}  \\ \hline
       \textbf{1} &  3.50 seconds \\ \hline
       \textbf{2} &  3.84 seconds \\ \hline
       \textbf{3} &  4.98 seconds \\ \hline
       \textbf{4} &  5.43 seconds \\ \hline
       \textbf{5} &  7.18 seconds \\ 
     \hline
    \end{tabular}
    }
   \vspace{-2mm}
  \end{center}
\end{table}

\textit{Overhead of increased number of IoT devices}: The results of this experiment are presented in Table \ref{tab:scalability1}. As can be seen, increasing the number of IoT devices that the LN gateway is serving resulted in an increase on the overall payment delays. Each Pi generated 100 key send payments all at once resulting in a burst of payments that need to be processed by the LN nodes running at the LN gateway. Thus, the increase in delays is not unusual and quite expected as the LN gateway communicated with each Pi more frequently for threshold signing operations. Additionally, the increase is not linear. When we check the results in Table \ref{tab:scalability2}, we observe that there is already a natural increase in delays when an LN node receives increasing number of transactions. This is because the LN protocol have to process all these HTLCs simultaneously which causes payment settling times to increase. Thus, we can understand that most of the increase in delays in Table \ref{tab:scalability1} are due to the LN overhead for processing the payments. The additional overhead coming from using our approach is minimal, which demonstrates that our approach can scale. Nevertheless, if certain delay requirements are to be met, additional LN gateways can be deployed to reduce the load depending on the number of IoT devices to be deployed.

\begin{table}[h]
%  \vspace{-5mm}
  \begin{center}
  \vspace{-3mm}
    \caption{Effect of Increasing the Number of Transactions on Payment Delays}
    \vspace{-6mm}
    \label{tab:scalability2}
    \resizebox{\linewidth}{!}{
    \begin{tabular}{|c|c|c|}
    \hline
       \textbf{Number of Tx}  &  \textbf{Threshold Delay} & \textbf{Baseline Delay} \\ \hline
       \textbf{100} & 3.50 seconds & 2.65 seconds  \\ \hline
       \textbf{200} & 5.87 seconds & 4.59 seconds \\ \hline
       \textbf{300} & 8.65 seconds & 6.96 seconds \\ \hline
       \textbf{400} & 10.53 seconds & 8.96 seconds \\
     \hline
    \end{tabular}
    }
   \vspace{-2mm}
  \end{center}
\end{table}

\textit{Overhead of increased number of payments}: 
The results of this experiments are shown in Table \ref{tab:scalability2}. Increasing the number of key send payments from a single sender resulted in increased delays for both threshold and baseline approaches. This clearly tells us that when an LN node processes more payments, some of the payments settle after the others resulting in a greater average delay. Thus, the increase in the delay is not relevant to our approach rather related to the LN's internal mechanisms. However, the threshold delays are still slightly higher than the baseline delays as expected which is due to the time spent on performing the threshold signing operations with the Pi.

\subsection{Energy Consumption}
\begin{figure}[h]
    \centering
    % \vspace{-3mm}
    \includegraphics[width=0.6\linewidth]{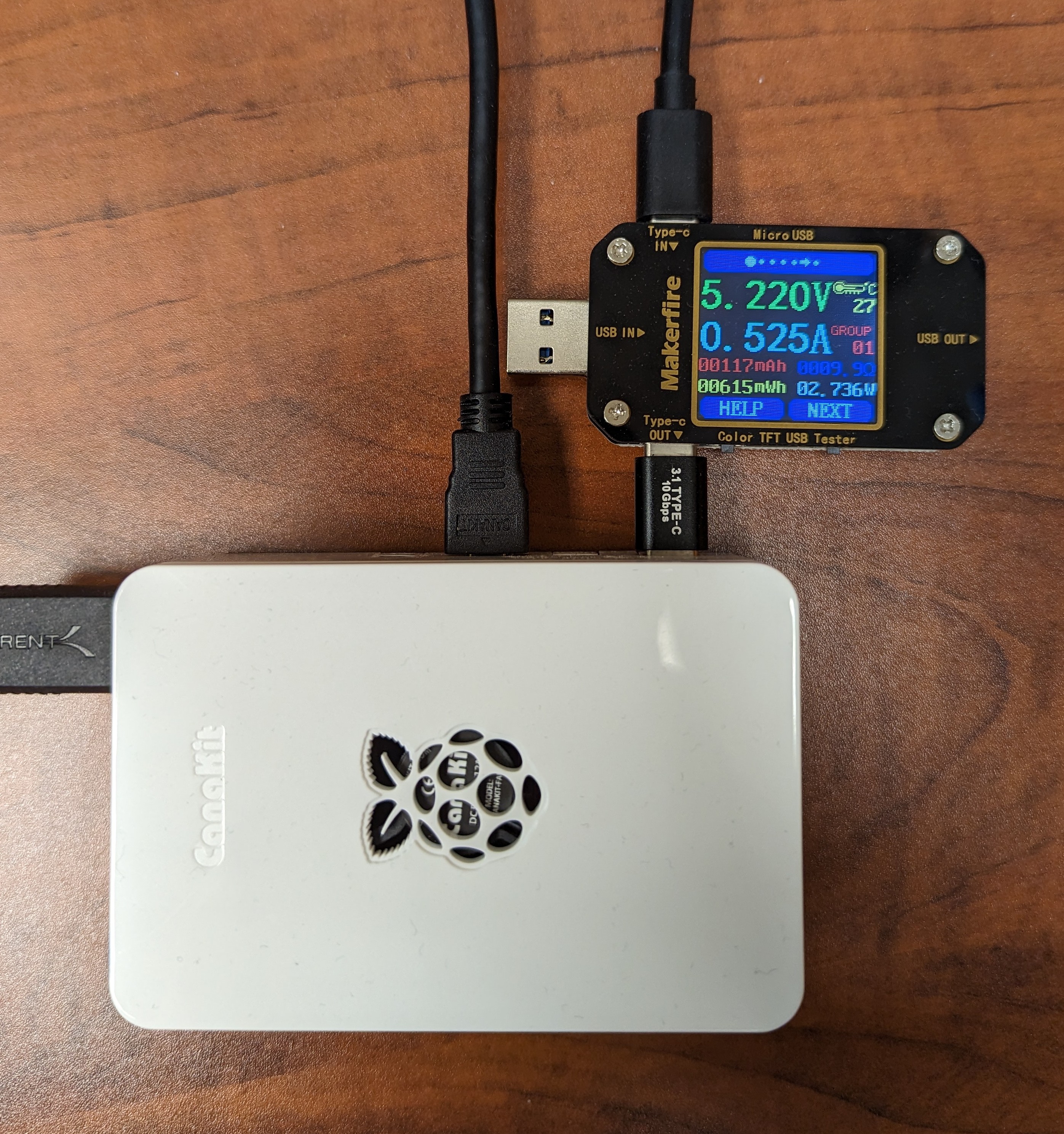}
    \vspace{-3mm}
    \caption{Our energy consumption setup for the Raspberry Pi using the MakerHawk USB Multimeter.}
    \vspace{-2mm}
    \label{fig:energy_setup}
\end{figure}

As many IoT devices are powered with batteries, it is important to also analyze their energy consumption. In this direction, we used a MakerHawk USB multimeter device which can accurately measure the power consumption of a Raspberry Pi in real-time. A photo of this setup is shown in Fig. \ref{fig:energy_setup}. We report the energy consumption of the Pi when it is idle and sending key send payments. We considered two ends of the spectrum: 100 and 400 transactions to see the difference. For the idle cases, we measured the energy consumption for the duration of time spent on sending the transactions. The results are shown in Table \ref{tab:energy}.

\begin{table}[h]
%  \vspace{-5mm}
  \begin{center}
  \vspace{-4mm}
    \caption{Energy Consumption of the IoT Device}
    \vspace{-3mm}
    \label{tab:energy}
    \resizebox{0.95\linewidth}{!}{
    \begin{tabular}{|c|c|c|}
    \hline
       \textbf{Number of Tx} & \textbf{Sending Payments} & \textbf{Idle} \\ \hline
       100 & 3.56 mWh & 3.03 mWh \\ \hline
       400 & 11.76 mWh & 10.13 mWh \\
     \hline
    \end{tabular}
    }
   \vspace{-3mm}
  \end{center}
\end{table}

According to these results, a single transaction only costs around 5 $\mu$Wh which indicates that the additional energy overhead that comes with our approach is minimal. Additionally, the overhead remains constant even if the number of transactions increase. Using battery solutions such as PiSugar\footnote{\url{https://www.pisugar.com/}}, Pis can run on battery around 10 hours non-stop. However, in our approach, we do not need the Pis to be running all the time. They can power-on only when they need to perform the threshold operations with the LN gateway and thus, can run on battery for days.

\subsection{Comparison with Other Methods}
\label{sec:comparison}

Since there are not any reliable comparison options, we chose to implement the state-of-the-art method for regular online payments which is the credit card payments. Specifically, we used Square's Web Payments SDK\footnote{\url{https://github.com/square/web-payments-quickstart}} to implement the standard credit card payments. A card payment utilizes 3D Secure (3DS) also known as Strong Customer Authentication (SCA). Square lets developers test credit card payments using their Sandbox servers. Thus, we installed this implementation into one of the Raspberry Pis and measured the delay to make card payments to their server. We tested the delays both with 3DS and without and present these results in Table \ref{tab:comparison}. The results clearly show that LN payments with our approach is much quicker than credit card payments even when 3DS is not enabled.

\begin{table}[h]
%  \vspace{-5mm}
  \begin{center}
  \vspace{-4mm}
    \caption{Comparison with Regular Credit Card Payments}
    \vspace{-4mm}
    \label{tab:comparison}
    \resizebox{0.95\linewidth}{!}{
    \begin{tabular}{|c|c|}
    \hline
       \textbf{Method} & \textbf{Payment Sending Time}  \\ \hline
       Credit Card Payment with SCA & 	$\approx$ 5 s \\ \hline 
       Credit Card Payment w/o SCA & 	$\approx$ 3 s  \\ \hline
       LNGate$^2$      &    1.01 s         \\
     \hline
    \end{tabular}
    }
   \vspace{-8mm}
  \end{center}
\end{table}

\section{Limitations}
\label{sec:limitations}

Our proposed protocol requires changes to the LN protocol. Specifically, we propose changes to LN's BOLT \#2 and commitment transactions. This will require ensuring that the updated LN software should be run on the LN gateway and bridge LN nodes in order to be able to support the IoT device transactions.

Additionally, our protocol requires the IoT device to be online for receiving payments. In an ideal setting, the IoT device should be able to receive a payment even when it is offline. It is important to note that this is also a problem for LN in general \cite{offlinelnnodes}.

\section{Conclusion}
\label{sec:conclusion}
In this paper, we proposed a secure and efficient protocol for enabling IoT devices to use Bitcoin's LN for sending and receiving payments. By introducing (2,2)-threshold scheme to LN and modifying LN's existing peer protocol for channel management and commitment transactions, a third peer (i.e., IoT device) was added to the LN channels. The purpose was to enable resource-constrained IoT devices that normally cannot interact with LN to interact with it and perform micro-payments with other users. IoT device's interactions with LN are achieved through an untrusted gateway node that has access to LN and thus can provide LN services in return for a fee. A game theoretic security analysis was provided to prove that the IoT device does not lose money when other channel peers attempt to cheat. Our evaluation results showed that the proposed protocol is scalable and enables IoT devices to use LN with negligible delay, cost, networking and energy consumption overheads.

% if have a single appendix:
%\appendix[Proof of the Zonklar Equations]
% or
%\appendix  % for no appendix heading
% do not use \section anymore after \appendix, only \section*
% is possibly needed

% use appendices with more than one appendix
% then use \section to start each appendix
% you must declare a \section before using any
% \subsection or using \label (\appendices by itself
% starts a section numbered zero.)
%

% \appendices
% \section{Proof of the First Zonklar Equation}
% Appendix one text goes here.

% % you can choose not to have a title for an appendix
% % if you want by leaving the argument blank
% \section{}
% Appendix two text goes here.

% use section* for acknowledgment
\ifCLASSOPTIONcompsoc
  % The Computer Society usually uses the plural form
  \section*{Acknowledgments}
  We would like to thank Christian Decker for helping us better understand how some LN protocols work. We also thank Ricardo Harrilal-Parchment for his help on creating the TCP/IP over BLE setup.
\else
  % regular IEEE prefers the singular form
  \section*{Acknowledgment}
\fi

% The authors would like to thank...

% Can use something like this to put references on a page
% by themselves when using endfloat and the captionsoff option.
\ifCLASSOPTIONcaptionsoff
  \newpage
\fi

% trigger a \newpage just before the given reference
% number - used to balance the columns on the last page
% adjust value as needed - may need to be readjusted if
% the document is modified later
%\IEEEtriggeratref{8}
% The "triggered" command can be changed if desired:
%\IEEEtriggercmd{\enlargethispage{-5in}}

% references section

% can use a bibliography generated by BibTeX as a .bbl file
% BibTeX documentation can be easily obtained at:
% http://mirror.ctan.org/biblio/bibtex/contrib/doc/
% The IEEEtran BibTeX style support page is at:
% http://www.michaelshell.org/tex/ieeetran/bibtex/
\bibliographystyle{IEEEtran}
% argument is your BibTeX string definitions and bibliography database(s)
\bibliography{IEEEabrv,references}
%
% <OR> manually copy in the resultant .bbl file
% set second argument of \begin to the number of references
% (used to reserve space for the reference number labels box)

% biography section
% 
% If you have an EPS/PDF photo (graphicx package needed) extra braces are
% needed around the contents of the optional argument to biography to prevent
% the LaTeX parser from getting confused when it sees the complicated
% \includegraphics command within an optional argument. (You could create
% your own custom macro containing the \includegraphics command to make things
% simpler here.)
%\begin{IEEEbiography}[{\includegraphics[width=1in,height=1.25in,clip,keepaspectratio]{mshell}}]{Michael Shell}
% or if you just want to reserve a space for a photo:

\begin{IEEEbiography}[{\includegraphics[width=1in,height=1.25in,clip,keepaspectratio]{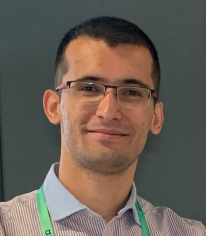}}]{Ahmet Kurt} received two B.S. degrees from Antalya Bilim University, Antalya, Turkey in 2018. He received the M.Sc. degree from the Department of Electrical and Computer Engineering, Florida International University, Miami, Florida in 2021. He is currently pursuing the Ph.D. degree in Electrical and Computer Engineering with the Florida International University, Miami, United States. His current research interests include
Bitcoin’s lightning network, Bitcoin and wireless networks.
\end{IEEEbiography}

\vspace{-10mm}

\begin{IEEEbiography}[{\includegraphics[width=1in,height=1.25in,clip,keepaspectratio]{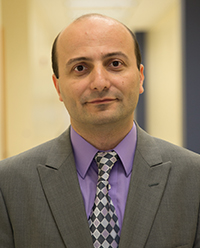}}]{Kemal Akkaya} 
% is a full professor in the Department of Electrical and Computer Engineering with a joint courtesy appointment in the School of Computer and Information Sciences at Florida International University. He received his PhD in Computer Science from University of Maryland Baltimore County in 2005 and joined the department of Computer Science at Southern Illinois University (SIU) as an assistant professor. Dr. Akkaya was an associate professor at SIU from 2011 to 2014. He was also a visiting professor at The George Washington University in Fall 2013. Dr. Akkaya leads the Advanced Wireless and Security Lab (ADWISE) in the ECE Dept. His current research interests include security and privacy, internet-of-things, and cyber-physical systems. Dr. Akkaya is a senior member of IEEE. He is the EiC of Springer Computer Science journal, area editor of Elsevier Ad Hoc Network Journal and serves on the editorial board of IEEE Communication Surveys and Tutorials. Dr. Akkaya was the General Chair of IEEE LCN 2018 and TPC Chair for IEEE ICC Smart Grid Communications. He has served as the guest editor for many journals and in the OC/TPC of many leading network/security conferences including IEEE ICC, Globecom, ICNP, INFOCOM, LCN and WCNC and ACM WiSec. He has published over 230 papers in peer reviewed journal and conferences. He has received ``Top Cited'' article award from Elsevier in 2010. 
received the Ph.D. degree in computer science from the University of Maryland, Baltimore, MD, USA, in 2005. He joined, as an Assistant Professor, the Department of Computer Science, Southern Illinois University Carbondale (SIU), Carbondale, IL, USA, where he was an Associate Professor from 2011 to 2014. He was also a Visiting Professor with George Washington University, Washington, DC, USA, in 2013. He is currently a Professor with the Department of Electrical and Computer Engineering, Florida International University, Miami, FL, USA. His current research interests include security and privacy, IoT and cyber-physical systems. He was the recipient of the Top Cited Article Award from Elsevier in 2010. He is currently an Area Editor for the Elsevier Ad Hoc Network journal, and is on the Editorial Board of the IEEE Communication surveys and tutorials. 
\end{IEEEbiography}

\vspace{-10mm}

\begin{IEEEbiography}[{\includegraphics[width=1in,height=1.25in,clip,keepaspectratio]{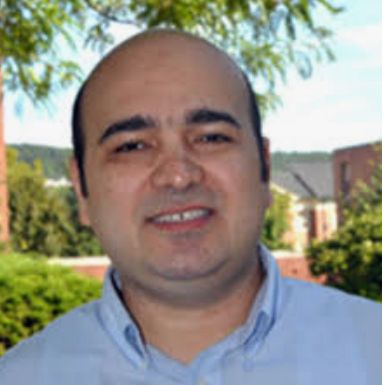}}]{Sabri Yilmaz} is an Associate Teaching Professor of Economics at Penn State-Harrisburg. He earned his Ph.D. degree in Economics from Southern Illinois University Carbondale in 2012. His main research is in the areas of applications of Game Theory and Economic Network Models. He has published in economics and interdisciplinary journals such as Journal of Real Estate Portfolio Management and Ad Hoc Networks.
\end{IEEEbiography}

\vspace{-10mm}

\begin{IEEEbiography}[{\includegraphics[width=1in,height=1.25in,clip,keepaspectratio]{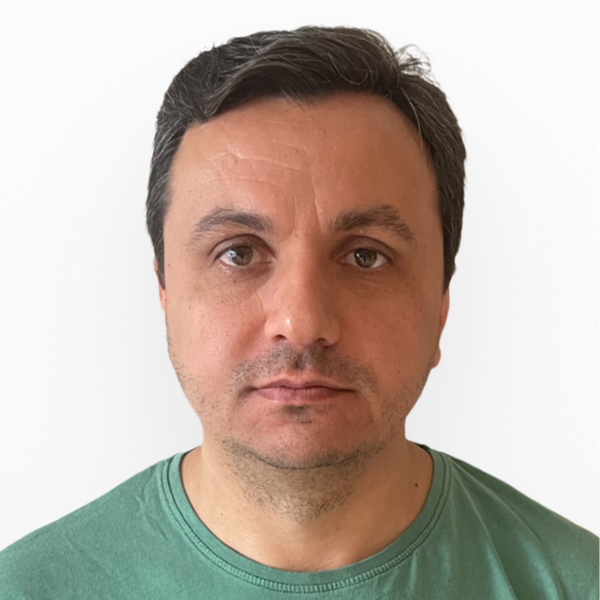}}]{Suat Mercan} is a Senior Software Engineer at Google. He was previously a Postdoctoral Researcher at Florida International University. Prior to that, he was an Assistant Professor at Zirve University in Turkey and American University of Middle East, Kuwait. He received his Ph.D. degree in Computer Science from the University of Nevada, Reno in 2011 and his M.S. degree in Electrical and Computer Engineering from the University of South Alabama in 2007. His main research interests are blockchain, payment channel and peer-to-peer networks, cybersecurity, digital forensics, and content delivery.
\end{IEEEbiography}

\vspace{-10mm}

\begin{IEEEbiography}[{\includegraphics[width=1in,height=1.25in,clip,keepaspectratio]{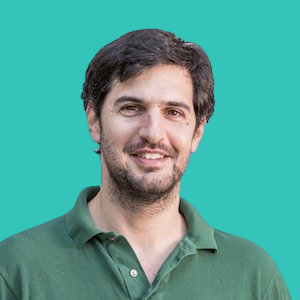}}]{Omer Shlomovits} is the previous co-founder of ZenGo, a mobile wallet based on multi-party computation. He led the cryptography research at the company. He is currently the CEO of Ingonyama, a semiconductor company specialized in Zero-Knowledge hardware accelerators.
\end{IEEEbiography}

\vspace{-10mm}

\begin{IEEEbiography}[{\includegraphics[width=1in,height=1.25in,clip,keepaspectratio]{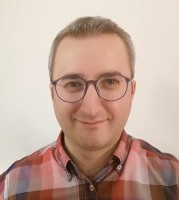}}]{Enes Erdin} is an Assistant Professor in the Department of Computer Science and Engineering at University of Central Arkansas, Conway. He conducts research in the areas of hardware security, blockchain technology, and cyber-physical systems. Erdin received a Ph.D. in Electrical and Computer Engineering from Florida International University, Miami where he was a NSF CyberCorps fellow.
\end{IEEEbiography}

% \vfill

% \begin{IEEEbiography}{Michael Shell}
% Biography text here.
% \end{IEEEbiography}

% % if you will not have a photo at all:
% \begin{IEEEbiographynophoto}{John Doe}
% Biography text here.
% \end{IEEEbiographynophoto}

% % insert where needed to balance the two columns on the last page with
% % biographies
% %\newpage

% \begin{IEEEbiographynophoto}{Jane Doe}
% Biography text here.
% \end{IEEEbiographynophoto}

% You can push biographies down or up by placing
% a \vfill before or after them. The appropriate
% use of \vfill depends on what kind of text is
% on the last page and whether or not the columns
% are being equalized.

%\vfill

% Can be used to pull up biographies so that the bottom of the last one
% is flush with the other column.
%\enlargethispage{-5in}

% that's all folks
\end{document}